# Potential solar axion signatures in X-ray observations with the XMM-Newton observatory

G.W. Fraser [1], A.M. Read [2*], S. Sembay [2], J.A. Carter [2] and E. Schyns [3]

[1] *Space Science and Instrumentation Group, Space Research Centre, Michael Atiyah Building, Department of Physics and Astronomy, University of Leicester, Leicester LE1 7RH, UK*
[2] *X-ray and Observational Astronomy Group, Department of Physics and Astronomy, University of Leicester, Leicester LE1 7RH, UK.*
[3] *PHOTONIS France S.A.S., avenue Roger Roncier, 19100 Brive, B.P. 520, 19106 BRIVE Cedex, France.*

## Abstract

The soft X-ray flux produced by solar axions in the Earth's magnetic field is evaluated in the context of ESA's XMM-Newton observatory. Recent calculations of the scattering of axion-conversion X-rays suggest that the sunward magnetosphere could be an observable source of 0.2-10 keV photons. For XMM-Newton, any conversion X-ray intensity will be seasonally modulated by virtue of the changing visibility of the sunward magnetic field region. A simple model of the geomagnetic field is combined with the ephemeris of XMM-Newton to predict the seasonal variation of the conversion X-ray intensity. This model is compared with stacked XMM-Newton blank sky datasets from which point sources have been systematically removed. Remarkably, a seasonally varying X-ray background signal is observed. The EPIC count rates are in the ratio of their X-ray grasps, indicating a non-instrumental, external photon origin, with significances of 11σ (pn), 4σ (MOS1) and 5σ (MOS2). After examining the distribution of the constituent observations spatially, temporally and in terms of the accepted representation of the cosmic X-ray background, we conclude that this variable signal is consistent with the conversion of solar axions in the Earth's magnetic field, assuming the resultant photons are not strictly forward-directed, and enter the field-of-view of XMM-Newton. The spectrum is consistent with a solar axion spectrum dominated by bremsstrahlung- and Compton-like processes, distinct from a Primakoff spectrum, i.e. axion-electron coupling dominates over axion-photon coupling and the peak of the axion spectrum is below 1 keV. A value of $2.2 \times 10^{-22}$ GeV$^{-1}$ is derived for the product of the axion-photon and axion-electron coupling constants, for an axion mass in the μeV range. Comparisons, e.g., with limits derived from white dwarf cooling may not be applicable, as these refer to axions in the ~0.01 eV range. Preliminary results are given of a search for axion-conversion X-ray lines, in particular the predicted narrow features due to silicon, sulphur and iron in the solar core, and the 14.4 keV transition line from $^{57}$Fe.

1. **Introduction**

The direct detection of dark matter has preoccupied Physics for over thirty years. Of the current candidate dark matter particles, axions – using the term indiscriminately to encompass the several families of weakly-interacting, light, neutral, spin-zero bosons - may be observable as a result of their mixing with photons in an external electromagnetic field, either astrophysical or laboratory-based (Asztalos et al. 2006; Raffelt et al. 2007).

**1.1 GECOSAX**
**1.1.1 Primakoff spectrum**

Extending the work of Di Lella and Zioutas (2003), Davoudiasl and Huber (hereafter DH) (2006, 2008) proposed viewing the axion-emitting solar core through the solid Earth with an X-ray telescope. Solar axions converting via the inverse Primakoff effect (Sikivie, 1983) in the nightside planetary magnetic field between the Earth and an orbiting spacecraft should give rise to an upwelling soft X-ray flux with a thermal spectrum peaking at 3 keV and with a mean energy ~4.2 keV. This observing mode – the Geomagnetic Conversion of Solar Axions into X-rays (GECOSAX) - excludes both the X-ray bright solar disk and the diffuse cosmic X-ray background (CXB) and so provides high sensitivity to the point-like axion-induced signal for an X-ray telescope in Low Earth Orbit (LEO). The 2-6 keV energy band where this signal is expected to have its maximum is relatively free of instrumental line features in contemporary solid-state focal plane detectors.

For a telescope pointing sunwards through the solid Earth, the conversion X-ray count rate in an energy interval $E, E+dE$ is given by:

$$N_x(E) = A \frac{dF_a}{dE} p dE \quad \text{photons/s} \quad \text{-(1)}$$

where: $A$ is the telescope effective area and $dF_a/dE$ is the solar axion flux at 1 AU. This flux has the expected differential form (DH 2006,2008):

$$\frac{dF_a}{dE} = 6.02 x 10^{10} g_{10}^2 E^{2.481} \exp(-E/1.205) = k_1 g_{10}^2 f(E) \quad \text{axions/cm}^2\text{/s/keV} \quad \text{-(2)}$$

with $E$ in keV; $g_{10}$ is the axion-photon coupling constant $g_{a\gamma}$ expressed in units of $10^{-10}$ GeV$^{-1}$ and $p$ is the axion-to-X-ray conversion probability, related to the magnetic field $B\perp$ at right angles to the direction of axion motion (i.e. perpendicular to the Sun-Earth line) and the conversion length $L$ by (Sikivie 1983; DH 2006, 2008) :

$$p = (\frac{E}{4(E^2 - m_a^2)^{1/2}})(g_{a\gamma} B_\perp L)^2 [\frac{2}{qL} \sin(qL/2)]^2 \quad \text{-(3a)}$$

which simplifies to :

$$p = (g_{a\gamma})^2 . B\perp^2 L^2/4 = k_2(g_{10})^2 B\perp^2 L^2 \quad \text{-(3b)}$$

in the limit $qL \rightarrow 0$. The parameter $q$ represents the longitudinal momentum difference between the massive axion and the massless X-ray photon, a difference which is taken up by a virtual photon and which in vacuum (and in natural units) has the form:

$$q = m_a^2 / 2E \quad \text{-(4)}$$

The ratio $\pi/q$ is interpreted as the oscillation length, $L_{osc}$, between axion and photon states. Setting the oscillation length equal to the size of the conversion field region $L$ (i.e. the satellite's orbital altitude in the DH analysis) defines the upper boundary of the axion mass range for which the conversion is coherent and to which the experiment is most sensitive.

In the GECOSAX geometry, observing from 600 km near-equatorial LEO, maximum sensitivity is for ~8 x $10^{-4}$ eV axions (DH 2006, 2008). The product $B_\perp L$ is ~ 3 x$10^{-5}$ T times 6 x$10^5$ m or 18 Tesla-metres, within a factor ~5 of the corresponding figure-of-merit for the CERN Axion Solar Telescope (CAST) helioscope, based on the 9.26m bore of an 9T superfluid-conducting electromagnet. According to DH, $k_2$~2.45 x $10^{-21}$. Neglecting the actual variation in focal plane detector quantum efficiency with X-ray energy $E$, and adopting $A$~800 cm$^2$ (Section 2.1, below) for the telescope collecting area along with $g_{10} = 1$ (below), the axion point source spectrum is initially represented by the curve labelled "Primakoff" in Fig. 1a. The final prediction of the GECOSAX model is a readily-detectable total 0.1-10 keV count rate of ~22 events in $10^5$ seconds, in a point source coincident, within a few arcminutes, with the centre of the Sun, but depending on the axion-photon coupling constant raised to the fourth power.

*Suzaku* X-ray Imaging Spectrometer (XIS) observations of the dark Earth (Katayama et al. 2005), X-ray spectra of the Sun in the 3-6 keV energy band from *RHESSI* (Hannah et al. 2007) and the X-ray Solar Monitor on the lunar probe *SMART-1* (Zioutas et al. 2006) have now set astrophysical upper limits to the Primakoff solar axion flux which, although they do not yet match the sensitivity of ground-based axion helioscopes such as CAST (Zioutas et al. 2007 ; Arik et al. 2009) and the (2.3m, 4T) Japanese Sumico (Inoue et al. 2008), call into question the basic assumptions of the DH model. The X-ray point source at the centre of the solar disc is not readily observed. The X-ray spectrum of the quiet Sun does not follow eq. 2, but keeps on rising below 1 keV (Zioutas et al. 2009).

The window of allowed axion masses, determined from ground-based searches and from observations of astrophysical objects including the Sun (Chelouche et al. 2009) is generally considered to be:

$10^{-6}$ eV < $m_a$ < $10^{-3}$ eV

while the axion-photon coupling constant $g_{a\gamma}$ is constrained, principally by CAST, to be less than or equal to $10^{-10}$ GeV$^{-1}$.

## 1.1.2 Bremsstrahlung and Compton spectra

Recent papers explore a more complex (and more detectable) multi-component form for the axion conversion X-ray spectrum. For example, Derbin et al. (2011, 2012, 2013) present numerical estimates of the axion production rates inside the Sun via two pathways additional to the Primakoff effect, both dependent on the axion-electron coupling constant, $g_{ae}$, rather than on the axion-photon coupling constant. For the Compton-like process mediated by electrons in the solar plasma ($\gamma + e^- \rightarrow e^- + a$), the analogue of eq. 2 is :

$$\frac{dF_a}{dE} = 1.33x10^{33} g_{ae}^2 E^{2.89} \exp(-0.774E) = k_3 g_{11}^2 f_c(E) \quad \text{axions/cm}^2\text{/s/keV} \qquad -(5)$$

while for the bremsstrahlung-like process ($e^- + Z \rightarrow e^- + Z + a$) :

$$\frac{dF_a}{dE} = 4.14x10^{35} g_{ae}^2 E^{0.89} \exp(-0.7E - 1.26E^{0.5}) = k_4 g_{11}^2 f_b(E) \quad \text{axions/cm}^2\text{/s/keV} \qquad -(6)$$

In both these equations, $g_{11}$ denotes the axion-electron coupling constant expressed in units of $10^{-11}$ and $E$ is again expressed in keV. Even more axion creation pathways (free-free emission, axion recombination and the M1 nuclear magnetic transition in $^{57}$Fe) have recently been evaluated by the CAST and EDELWEISS II collaborations (Armengaud et al. 2013; Barth et al. 2013). The $^{57}$Fe channel gives rise to a narrow 14.4 keV axion line whose intensity (see Section 4.4.4) depends on the square of the effective axion-nucleon coupling constant, $g_{aN}$. Axion-ion interaction channels may also result (Redondo, 2013) in line complexes close to, but not perfectly coincident with, the K-shell energies of either neutral matter or of the hydrogen- or helium-like ions of silicon, sulphur and iron.

Fig. 1a shows the energy dependence of the three spectral components described in Eqs. 2, 5 and 6, computed for $g_{10} = g_{11} = 1$, consistent with most current experiments (but about one order of magnitude greater, in terms of their product, than the joint estimate given by Barth et al. (2013)). The bremsstrahlung component dominates over the Compton component and over the Primakoff signal and peaks at a much lower energy than either of them – at about 0.6 keV, rather than 3 keV. In the limit $g_{10} = g_{11} = 1$, the expected axion conversion spectrum is softer than the canonical extragalactic cosmic X-ray background (CXB), whose photon index equals 1.4. For $g_{11}$ values less than unity, and $g_{10}$ maintained constant, the local photon index at 6 keV decreases from ~2.4 to ~2.2 and the spectrum hardens slightly (see Fig. 1b).

The new, multi-component models appear fit for purpose when confronted with X-ray spectra of the quiet Sun (Zioutas et al. 2009) but have not previously been tested against observations of cosmic sources.

## 1.2 Practical GECOSAX
### 1.2.1 XMM-Newton

Of all the instruments considered by DH, the European Photon Imaging Camera (EPIC) on board the European Space Agency's XMM-Newton X-ray observatory (Jansen et al. 2001), launched on 10[th] December 1999, provides by far the largest product of effective area $A$ (~800 cm$^2$ at 4 keV for the EPIC pn channel) and accumulated observation time $t$ (~3 x10$^8$ s) for the potential detection of axion conversion X-rays. XMM's effective area-observing time-field of view "triple product" is now, following 2010-13 mission selections by ESA and NASA, unlikely to be exceeded by any future X-ray observatory (Smith et al. 2010) for the next twenty years.

XMM-Newton operations, however, exclude pointing at the Sun and at the Earth[1]. Assuming that the X-rays produced when axions convert in a transverse magnetic field propagate in the direction of the original particles, it would appear unlikely that the XMM-Newton archive contains explicit information on the solar axion observables $dF_a/dE$, $m_a$, $g_{ae}$ and $g_{a\gamma}$. EPIC, like other X-ray instruments, has of course been extensively used to search for the spectral and/or spatial signatures of dark matter in the lines-of-sight to distant cosmic objects (Boyarski et al. 2012, 2014; Bulbul et al. 2014).

An X-ray observatory whose pointing direction is on average randomly orientated to the Sun-Earth line can, in this one-dimensional picture, detect only those conversion X-ray photons, produced in the Earth's magnetic field, which subsequently have been elastically scattered, on average, through a right angle. This observing geometry – a kind of *"orthogonal GECOSAX"* configuration for light solar axions or other similar particles – has one very significant disadvantage and one equally-significant advantage compared to the original ideal geometry proposed by DH. The disadvantage is the need to detect a faint extended source filling the field-of-view rather than a Sun-centred point source. However, for a spacecraft such as XMM-Newton in an elliptical High Earth Orbit (HEO), any conversion flux that enters the XMM-Newton field-of-view would be highly modulated with a period of one year, compared with the GECOSAX conversion X-ray flux, whose seasonal depth-of-modulation would be expected only at a level of ~6%, the combined result of the eccentricity of the Earth's orbit and the inverse square law. This high degree of modulation would arise from the fact that the direction of the Earth's magnetotail is always away from the Sun, while the satellite orbit is "fixed" in inertial space (see Figs. 2a,b and 3a,b). An X-ray telescope in HEO can therefore sample volume measures of the square of the transverse magnetic field $B\perp^2 = B_y^2 + B_z^2$ - determining the rate of X-ray production (see eq. 3b) - which vary considerably from season to season.

---

[1] This is true for almost all X-ray astronomy satellites; the exception is the Franco-Chinese SVOM Gamma-ray Burst (GRB) mission, which will view the night side of the Earth for part of every orbit (Götz et al. 2009).

The difference X-ray spectrum obtained by subtracting "Winter" (i.e. the season of expected minimum signal) from "Summer" (expected maximum signal) should then follow a profile $F(E) = f(E) + f_c(E) + f_b(E)$ from Figs. 1a,b, and tend to zero count rate at energies above ~6 keV and at energies below ~0.5 keV. The well-defined spectral shape of the GECOSAX signal is now rendered uncertain in proportion to the ratio $(g_{11}/g_{10})^2$.

### 1.2.2 Organisation of Paper

Since the inefficiency[2] of the elastic scattering process is, in fact, unlikely to be compensated by any increase in magnetic volume compared to the original DH geometry (Section 2.2), this work was prompted by recent hints that the motion of axions and their conversion X-rays need not be perfectly co-linear in inhomogeneous magnetic fields (Guendelman 2008; Guendelman et al. 2010, 2012). These papers investigate the conversion probability $p$ due to "axion splitting" in a number of ideal magnetic field geometries (infinitely thin solenoid, square well, Gaussian and δ-function) but not yet for the desired dipole approximation to the geomagnetic field. It is thought here that isotropic scattering axion-to-photon conversion probabilities can attain values of the same order as for purely co-linear scattering, especially when the axion mass and the energy corresponding to the plasma frequency are equal, and the conversion probability shows a resonance and increases sharply (Guendelman et al. 2010).

Furthermore, the inverse Compton effect ($e^- + a \rightarrow \gamma + e^-$) should now be considered as a potential axion-to-photon conversion mechanism alongside the (inverse) Primakoff effect. If inverse Compton conversion is indeed significant, the geomagnetic dipole field remains important, but only indirectly, in that it then acts as a proxy for the number density of the contained charged particles that scatter the conversion X-rays. Additionally, inside a magnetic field, the Primakoff effect can give rise to axion-to-photon conversion and also photon-to-axion back-conversion as soon as the photons start appearing (Zioutas et al. 2006), leading to a mix of axions and photons. Therefore, although the research along these different lines of enquiry is still in its early stages, we postulate that there may well be various mechanisms for conversion X-ray photons to enter the XMM-Newton field-of-view (see Figs. 2a,b).

Section 2 therefore explores, on a phenomenological basis, the observational consequences of replacing pure forward or backward scattering (Sikivie 1983) with isotropic (Guendelman et al. 2010) X-ray propagation, post conversion, of a multi-component axion spectrum.

Section 3 of this paper reports the results of a search for a "solar axion effect" in the 2-6 keV cosmic X-ray background (CXB) observed by the XMM-Newton EPIC pn and MOS cameras.

In fact, the XMM-Newton EPIC pn background is well known to be seasonally modulated, with repeating winter minima and summer maxima (Rodriguez-Pascual & Gonzalex-Riestra 2008) especially for those phases of the orbit when the spacecraft is emerging from, or about to enter, the trapped radiation belts. The ~1-10 cm$^{-2}$s$^{-1}$ magnitude of this high energy particle-induced signal is much greater than the representative conversion X-ray intensities calculated in Section 1.1.2 – but is also easily discriminated on the grounds of event size and morphology. A more likely limitation to the detection of an axion-related signal is the distinct soft proton (SP) population, with energies less than 500 keV, which is present in sporadic clouds throughout the magnetosphere. The Appendix attempts to construct a physical model of signal generation by SPs reaching the EPIC focal planes after being focused or scattered through the XMM-Newton X-ray optics. A large number of studies with precisely this aim have been published since the launch of XMM-Newton.

---

[2] The electron number density in the magnetotail/ plasma sheet is ~0.5 cm$^{-3}$ and the Thomson cross-section is $6.6 \times 10^{-25}$ cm$^2$, so that the probability that a photon would undergo elastic scattering in a distance $L \sim R_E$ is only of order ~2 x $10^{-16}$. $R_E$ is the radius of the Earth, 6370 km.

## 2. Axion-conversion X-ray signals in the XMM-Newton orbit
## 2.1 Coordinates and Fields

The origin of Geocentric Solar Ecliptic (GSE) coordinates is at the centre of the Earth and the Sun is in the +x direction. The Earth's magnetic field has components ($B_x$, $B_y$, $B_z$). The XMM-Newton spacecraft's instantaneous position is ($X_s$, $Y_s$, $Z_s$). An XMM-Newton mission ephemeris in GSE coordinates has been generated from NORAD two-line elements[3]. The orbital period at beginning of mission was 48 hr or 1.728 x 10$^5$ s and the apogee was 116,000 km in the southern hemisphere (see Figs. 2a,b and 3a,b). The initial inclination was 40º and the initial perigee, 7365 km.

Observations are made only when the spacecraft is above the Earth's radiation belts, above 40,000 km or 6.27 $R_E$. Because of solar panel constraints, no observations are made further than 20 degrees from the perpendicular to the Sun-Earth line; the Earth limb avoidance angle is 42º. A final observing constraint, related to Moon avoidance, is not implemented in our model described below.

According to eq. 3b, any axion-related X-ray signal depends on the square of $L$, the length scale of the magnetic field, and on the square of the transverse field strength, $B_\perp$. The geometric relationships of the observational lines-of-sight relative to the solid Earth and to the hypothetical axion flux do not change with time. Thus, in order to estimate the relative seasonal variation of the axion-induced X-ray signal, one evaluates, for the ensemble of all allowed pointing directions, a representative value of $B_\perp^2$ along the line-of-sight intercepting the Earth's magnetic field, repeating the calculation appropriately as the orbital geometry evolves from that of Fig. 2a to that of Fig. 2b.

The analytical model of Luhmann and Friesen (1979) was used to represent the geomagnetic field in preference to the definitive (but more computationally intensive) International Geomagnetic Reference Field (IRGF)[4], whose formal validity extends only to ~6.6 $R_E$. This steady-state magnetic field, without tilt, bow shock or explicit representation of the ~5nT average interplanetary magnetic field (IMF), is the sum of a dipole field and a current sheet field:

$$B_r = -2M cos\theta \, r^{-3} + B_T sin\theta \, cos\varphi \qquad \text{-(7a)}$$

$$B_\theta = -M sin\theta \, r^{-3} + B_T cos\theta \, cos\varphi \qquad \text{-(7b)}$$

$$B_\varphi = -B_T sin\varphi \qquad \text{-(7c)}$$

Here, $\theta$ is the co-latitude, $\varphi$ is the magnetic longitude measured from the –x (midnight) axis and $M$ is the dipole field strength of the Earth (~10$^2$ T.m). At large radial distances $r$, the field strength in the magnetotail is close in magnitude to the value $B_T$ (~15 nT). In the model equations, the $B_T$ term changes sign from positive (representing a sunward-pointing field in the northern hemisphere) to negative (an anti-sunward field) as $z$ changes sign. Eqs. 7a-c are claimed to be valid out to ~13.3 Earth radii (Luhmann & Friesen 1979), encompassing the larger part of the XMM-Newton orbit. Fig. 4 shows the sketch form of $B\perp^2$ in the noon-midnight plane.

---
[3] Two-line orbital elements available from: http://www.celestek.com/NORAD/elements.science.txt
[4] http://www.geomag.bgs.ac.uk

## 2.2   Calculations

For any given XMM-Newton revolution, random observational lines-of-sight were constructed from the instantaneous spacecraft position ($X_s$, $Y_s$, $Z_s$) at time intervals of 900s (the time step size of the orbital description). The actual length of the average EPIC observation is 23 ks (Carter & Read 2007), so that the model telescope boresight position changes many times during any real observation period. At spatial intervals $ds$ along that line-of-sight, the square of the local transverse magnetic field $B\perp^2 = B_y^2 + B_z^2$ was computed from the Luhmann and Friesen (1979) model. The length of the line-of-sight ($N_{max}.ds$) was chosen to encompass the entire magnetic system for all observing directions, while $ds$ was made small enough to represent the rapid variations of the near-Earth dipole field. In what follows, $ds = 0.035\ R_E$ and $N_{max} = 1000$; the results are not sensitive to the exact values of $ds$ and $N_{max}$.

For each individual pointing, we must account for the contradictory geometric effects along the line-of-sight of the expanding telescope field-of-view and the diminishing solid angle subtended by the telescope aperture which receives the "returning", isotropically-emitted, axion-related X-ray flux. Suppose each XMM-Newton telescope (there are three identical mirror modules) has an aperture $d\Omega$. Let the distance from the spacecraft along the pointing direction (defined by the unit vector ($a_1$, $b_1$, $c_1$)) be $s$. Then the intercepted axion flux is related to the size of the projected field-of-view of the telescope in the GSE ($y$, $z$) plane, which is :

$$\sim d\Omega s^2 a_1 \qquad\qquad -(8)$$

while the probability of a conversion X-ray originating in that field of view and subsequently entering the telescope is $A/(4\pi s^2)$. The product of these terms is independent of $s$ and a proxy for the axion-derived X-ray signal is simply:

$$S = \sum_1^{N\max} B_\perp^2 \qquad\qquad -(9)$$

whereas $S/N_{max}$ is a measure of the average field intensity encountered in the pointing direction. The closure of the summation $S$ follows from the rapid fall-off in field strength with distance from the centre of the Earth, but is likely to be inexact because of the simplistic representation of the field boundary with the Interplanetary Magnetic Field.

Figs. 5a,b show the calculated seasonal variation of the $S$-parameter at the beginning of the XMM-Newton mission and near its current half-way point, respectively. For the first day of every month, a single characteristic revolution is "flown" ten times, giving a total of over 1000 individual valid pointings. The monthly signals can be finally combined into four equal spacecraft seasons, A1 to A4, broadly equivalent to the true northern hemisphere seasons Winter, Spring, Summer and Autumn (see Section 3.1.3).

In Fig. 5a, the predicted axion conversion signal is, as expected, greater in Summer (A3) than in Winter (A1). This dependence of the axion signal on time of year is the opposite of that expected from the eccentricity of the Earth's orbit and the inverse square law, which requires a minimum signal at the Summer solstice. Comparing Fig. 5a with Fig. 5b, there clearly exist more complex $S$-profiles, depending on the projection of the XMM-Newton orbit onto the GSE (*x, y*) ecliptic plane. This projection is not always balanced about the Sun-Earth line, the GSE x-axis (see Fig. 3b). The spacecraft may have better "visibility" of the sunward high-field region in October than in April. The calculated amplitude of the seasonal variation in the signal *S*, after accounting for the eccentricity of the Earth's orbit (as in Fig. 5a), is by a factor ~2.

The field/orbit model also predicts that observations of X-ray sources lying north of the ecliptic – observations looking through the bulk of the magnetosphere - should in general feature a larger axion conversion signal than southward-directed pointings, but with the difference between hemispheres varying strongly with time of year, indeed changing sign in Autumn in the test case illustrated in Fig. 6.

## 2.3 Constraints

There are two conditions that any candidate axion conversion X-ray signal must satisfy:

### 2.3.1 Consistency with past CXB observations

For consistency with previous observations, the X-ray spectrum averaged over all four spacecraft seasons A1 - A4 should conform to accepted models of the isotropic extragalactic CXB, (80±6)% of which is currently believed to be due to unresolved point sources, mostly Active Galactic Nuclei (AGN). In the broad 3-60 keV band, the accepted form of the CXB spectrum is a cut-off power law (Morreti et al. 2009):

$$dF_x/dE = 7.877 \cdot E^{-1.29} \exp(-E/41.13) \quad \text{photons/cm}^2\text{/s/keV/sr} \quad\quad -(10a)$$

A pure power law alternative, which fits the 2-6 X-ray background observed (for example) by the Suzaku X-ray Imaging Spectrometer (XIS), a CCD camera, is (Katayama et al. 2005):

$$dF_x/dE = 9.7 E^{-1.4} \quad \text{photons/cm}^2\text{/s/keV/sr} \quad\quad -(10b)$$

The latter function, with small changes to the normalisation and photon index (see Table 1), describes the extragalactic CXB spectra previously derived from the XMM-Newton EPIC cameras themselves.

### 2.3.2 Consistency between cameras

The simultaneous observation of each X-ray field by the multiple XMM-Newton telescopes provides an internal consistency check for any potential axion-related signal.

Three identical, co-aligned telescopes, each made up of 58 gold-coated replicated nickel shells, focus X-rays onto five silicon-based focal plane arrays[5]. The EPIC pn CCD camera (36 cm$^2$ total detector area, with 10% outside the 30 arcminute diameter telescope field-of-view) accepts all the focused X-rays from one complete telescope. The remaining two telescopes are divided between the EPIC MOS1 and MOS2 cameras (each one consisting of seven 2.4 x 2.4 cm$^2$ conventional metal-oxide-silicon front-illuminated CCDs, constituting 40.3 cm$^2$ total detector area) and the nine back-illuminated CCDs reading out each of the dispersive Reflection Grating Spectrometers (RGS1 and RGS2). XMM-Newton thus produces five independent measurements of the combined X-ray and non-X-ray background for every pointed observation. While the cosmic ray particle-induced background of each camera is determined by the materials and exact geometry of its internal construction, the count rates from a truly diffuse X-ray source must be ordered solely according to instrumental X-ray grasp *G* (i.e. the effective area – aperture product). That is:

---

[5] The XMM-Newton User Handbook is online at: http://xmm.esac.esa.int/external

$$G_{pn} > G_{MOS} >> G_{RGS}.$$

The ratio $G_{pn}/G_{MOS}$ is ~ 3.0:1 just below the Au $M_V$ absorption edge in the mirror response function at 2.24 keV and ~ 3.2:1 at 3 keV, rising to 5.0:1 above 5 keV due to the roll-off of the quantum efficiency of the MOS CCDs. The inter-calibration of the EPIC pn and MOS cameras is discussed, for example, by Mateos et al. (2009). The RGS registers dispersed 0.3 - 2 keV X-rays[6]. Its background is dominated by soft protons and cosmic rays. Its susceptibility to the first of these sources makes the RGS, in principle, an informative soft proton "monitor" (see Rodriguez-Pascual & Gonzalez-Riestra (2008) and the Appendix).

## 3. X-ray observations
### 3.1 Analysis methods
#### 3.1.1 Energy band

The XMM-Newton EPIC pn and EPIC MOS database[7] 2000-2012 was analysed using extensions of the Blank Sky protocols of Carter & Read (2007). Only Full-Frame (not Extended Full Frame, windowed or timing mode) data was included. Concentrating on energies above 2 keV and below 6 keV allowed all three bandpass filters (Thin, Medium and Thick; see Appendix) to be analysed together, while effectively excluding sources of soft X-ray diffuse emission such as the Local Bubble and the Warm Hot Intergalactic Medium (WHIM). These sources are reviewed in detail by Galeazzi et al. (2011). The galactic plane ($|b_2|<10°$) itself was completely excluded from the analysis in order to avoid regions of bright diffuse hard galactic emission with energies extending above 2 keV.

Immediately below the 2 keV energy threshold lie the strong instrumental self-fluorescence lines of silicon (1.74 keV – MOS cameras only) and aluminium (1.49 keV). Above 6 keV, there are $K_\alpha$ lines due to Fe (6.4 keV) in both pn and MOS cameras and Ni (7.3 keV), Cu (8.04 keV) and Zn (9 keV) in EPIC pn only. Au L shell emission is evident in EPIC MOS at 9.7 and 11.4 keV – implying the inevitable presence of gold M-shell emission at 2.122 ($M_\alpha$) and 2.203 ($M_\beta$) keV and up to a series limit of ~3 keV. The only other significant line features expected within the 2-6 keV band are the $K_\alpha$ lines of chromium (5.4 keV; pn and MOS) and manganese (5.9 keV; MOS only).

#### 3.1.2 Data selection

Calibrated event files were created using the standard XMM-Newton Science Analysis System SAS (v11.0) epchain/emchain analysis tasks for all the full-frame datasets in the 2000 - 2012 EPIC database, utilising the most up-to date calibration available at the time. Single- and double-pixel events were used for EPIC pn. Singles, doubles, triples and quadruples were used for MOS. Standard PPS (pipeline processing system) source lists were used to remove all the detected sources to large radii. Datasets were then visually inspected. Those containing single-reflection arcs from out-of-field bright sources or unusual chip-to-chip or quadrant-to-quadrant variations were rejected, as were those containing the wings of very bright removed sources, confused source extraction regions, large diffuse sources, residual out-of-time features, no events, and where sources appeared in "out of field-of-view, but open to the sky" gaps in the instrument housing (EPIC MOS only).

Each event file was then filtered further to remove periods of high background 'flaring', due to soft protons (see Appendix) entering through the telescope aperture. Holes left by the removal of point sources were then filled in or 'ghosted' on an event-by-event basis and the resultant files visually inspected once more. On average, 80±10 sources were removed from each field.

---

[6] The RGS's high wavelength resolution over this limited energy range has been utilised, however, in a search for another dark matter candidate, the sterile neutrino (Boyarski et al. 2007).

[7] http://xmm2.esac.esa.int/external/xmm

The files then underwent a second, more thorough, removal of soft proton contamination. A 2-12 keV light curve (single events, FLAG=0, 100s binning) was extracted from a centred annulus with inner and outer radii 15 and 600 arcseconds, respectively. Good retained times were defined via Gaussian clipping (Snowden et al. 2004) to be those lying less than 3.3 sigma above the peak of the count rate histogram. These time periods were extracted to form a working event list.

An inspection of all the selected light curves and histograms was made and files showing any unusual deviations from a "clean" Gaussian histogram were removed. Often, these were low-exposure observations. All observations with exposure times (after SP cleaning) less than 2000s were therefore removed from consideration.

For each of the final files, the level of the residual soft proton contamination was checked via the method of De Luca & Molendi (2004). High-energy spectra were formed, corresponding to in- and out-of-field-of-view (in-FOV and out-FOV) regions of the focal plane. The in-FOV region extended from a radius of 10 arcminutes out to the edge of the FOV. The out-FOV region consisted of all of the detector area beyond the edge of the FOV, not open to the sky. For EPIC pn, the 9-12 keV band, assumed in previous analyses to contain only high-energy particle-induced events in the out-FOV zone, and to be free from contamination from instrumental Cu and Ni fluorescence lines, was used. For EPIC MOS, the 10-11.2 keV band (De Luca & Molendi 2004), free from instrumental Au L-shell lines, was employed.

The ratio $R$ could then be formed of the density of events within the optical focal plane – containing now only residual soft protons, a small number (given the high energy band selected) of focused X-rays and cosmic ray events – to the density of events in the marginal detector areas "out-FOV" containing (ideally) only cosmic ray events, but known in practice (Lumb et al. 2002) to include 7% singly-reflected X-rays. A 'best file' list with minimised soft proton contamination resulted from the restrictive (De Luca & Molendi (2004)) choice $R < 1.3$.

Then, for each best file, the desired 2-6 keV spectrum – but still not corrected for cosmic ray events - was formed from the full telescope FOV, a centrally-located circle of 800 arcsecond radius.

### 3.1.3 Charged particle background correction

In both EPIC cameras, but not in RGS, the optical path to the focal plane can be interrupted by a ~mm thick aluminium filter and Filter Wheel Closed (FWC) data accumulated[8]. The cosmic ray contribution to each of the best files can be estimated by comparing the high-energy (i.e. 9-12 keV for EPIC pn or 10-11.2 keV for EPIC MOS), out-FOV spectrum with that extracted from the appropriate FWC file.

A number of methods to account for cosmic rays using the FWC spectra were investigated. The baseline method used the FWC file closest in time to the observation. This could prove unsatisfactory, since successive FWC files are often quite widely separated in time – sometimes by as much as six months. The second method involved splitting the FWC data into (three) subsets of varying FWC count rate, and then using the subset most appropriate for the out-FOV count rate of the target file. For all three EPIC cameras, this method yielded essentially the same results (with poorer FWC counting statistics) since, when using the full FWC datasets, the shape of the FWC spectrum did not change with count rate. The full stacked FWC datasets were used for background subtraction in what follows.

---

8    The full sequence of FWC observations for EPIC pn can be found at:
http://xmm2.esac.esa.int/external/xmm_sw_cal/background/filter_closed/pn/FF/rate_vs_time_FF_2013_v1_norad.png. The general increase in FWC count rate from 5 s$^{-1}$ in 2000 to 10 s$^{-1}$ in 2010 corresponds not only to a complete solar cycle, but also follows the rising fraction of the 48 h orbit that XMM-Newton spends inside the radiation belts, according to Rosenquist et al. (2002). Count rate "spikes" at revolutions 822, 895 and 1383 appear to correspond to calculated spacecraft crossings of the Earth's anti-sunward plasma sheet (Rosenquist et al. 2002). Spectrally, the EPIC pn FWC particle background measured at beginning of mission decreased smoothly from 0.2 counts s$^{-1}$ keV$^{-1}$ at 2 keV to 0.1 counts s$^{-1}$ keV$^{-1}$ at 12 keV.

Each of the best files was assigned to one of the four "spacecraft seasons" denoted A1, A2, A3 and A4, rather than to a true (calendar) season. At the beginning of the XMM-Newton mission, the spacecraft "winter solstice" – when the orbit apogee lies on the negative GSE x-axis, actually occurred in mid-January (see Fig. 3a) so that spacecraft season Winter/A1 coincided with the calendar months December to February. By 2010, the spacecraft seasons were shorter (43 revolutions each, rather than 46) and A1 - centred throughout the mission on the spacecraft winter solstice – had its midpoint in September. The evolution of the extrema of the XMM-Newton orbit in relation to the spacecraft seasons is shown in Fig. 7. The use of spacecraft seasons in the analysis, rather than true seasons, avoids the long-term blurring of the geometries of Fig. 2a and Fig. 2b and maximises sensitivity to a modulated signal.

Figs. 8a,b show the long-term evolution of the FWC scaling factors for EPIC pn and EPIC MOS2, respectively. The similarity of the curves for the four spacecraft seasons indicates that there is no bias, in either EPIC pn or EPIC MOS, in the FWC background correction process towards any particular season.

The final EPIC pn data products were four stacked X-ray spectra, one for each spacecraft season, integrated over a twelve-year period, more than one nominal solar cycle, beginning a year before the maximum of Cycle 23 and culminating close to the maximum of Cycle 24 in December 2012. Each stacked seasonal spectrum had associated with it a correctly scaled FWC spectrum, describing the instrumental background.

Instrument response files were created using standard SAS tools. The instrument effective area file was calculated assuming the source flux to be extended and flat with no intrinsic spatial structure.

The EPIC pn analysis was repeated for the EPIC MOS2 camera for all revolutions and for EPIC MOS1 up to revolution 961 (March 2005), after which the MOS1 response and its FWC scaling are complicated by the loss of CCD6 to a micrometeoroid strike.

Apart from the final segregation of the files according to season, the data reduction methods follow standard procedures for processing diffuse emission seen by XMM-Newton (Carter & Read 2007). Nothing in the data reduction procedure presumes any feature of the solar axion conversion model described in Section 1, The rigour of the data screening process is indicated by the fact that only 17% of the available files were retained and only ~6% of the possible exposure time.

### 3.1.4 Diagnostics

The distribution of the selected observations by exposure time, position on the sky and by spacecraft season are summarised in Table 2 and illustrated in Figs. 9-13[9].

Table 2 indicates a preponderance of A1 "best" files compared to the other three seasons for EPIC pn and (to a lesser extent) for EPIC MOS. The only readily apparent result of this selection effect is the higher statistical precision of the A1. Table 2 also appears to show that, even before correction for residual cosmic rays, the selected events are consistent with an X-ray dominated signal, since the "average background" count rates in the EPIC cameras are approximately in proportion to their X-ray grasps (Section 2.3.2). However, such proportionality is also a property of penetrating high energy cosmic rays in a detector heavily shielded on five sides.

---

[9] The following colour scheme is adopted in all subsequent histograms and spectra: spacecraft season A1 (black), A2 (red), A3 (green) and A4 (blue).

Fig. 9a,b show the distributions of exposure times for the best EPIC pn and EPIC MOS2 files, resolved into seasons. The constancy of mean exposure time from season to season for both cameras implies a seasonally independent flux distribution for the removed point sources in each camera. Figs. 10a,b confirm that this flux distribution is indeed observed.

Figs. 11a,b show how the balance of observations between the four spacecraft seasons evolved with mission elapsed time.

Figs. 12a,b show the sky distributions of the A1-A4 observations for EPIC pn and EPIC MOS2, respectively. There is no preferential concentration of the selected best EPIC fields towards the galactic plane or poles, or indeed towards any known large-scale massive structures such as the Virgo /Coma or Centaurus regions (Loewenstein & Kusenko 2012). There are, in fact, fewer A3 observations than A1 observations at mid galactic latitudes, so that any observed "summer excess" is unlikely to be due to residual contamination by any emission extending out from the galactic plane.

Figs. 13a,b show the frequency distributions, for EPIC pn and EPIC MOS2 respectively, of the values of the "flux in, flux out" ratio $R$. There is no significant difference between the seasons in terms of $R$-value, for either camera. The detailed shape of the distribution differs between EPIC pn and EPIC MOS because of their different out-of-field detector areas.

## 3.2    Results: EPIC pn

The EPIC pn in-orbit background has been very extensively studied (Lumb et al. 2002; Rodriguez-Pascual & Gonzalez-Riestra 2008; Snowden Collier & Kuntz 2004). The non-X-ray component has been simulated using the Monte Carlo package GEANT4 by Tenzer et al. (2008). The laboratory background of a pn-CCD camera (actually, the focal plane sensor for CAST) is described by Kuster et al. (2005).

In the past, EPIC MOS has been generally preferred for studies of the extragalactic CXB, because of:

(a)     the imperfectly-characterised charge transfer efficiency in individual pn CCDs
(b)     the factor ~2 higher charged particle background in EPIC pn
(c)     the relatively small out-of-field-of-view pn detector area for the determination of the cosmic ray flux (Section 3.1.3).

Here, we are concerned with possible seasonal differences in an already faint diffuse signal, and the larger photon grasp of the EPIC pn camera is the key parameter.

Fig. 14a shows the individual 2-6 keV X-ray spectra for spacecraft seasons A1-A4, integrated from 2000-12. The Winter spectrum A1 clearly lies below the other three, while A4 (Autumn) is significantly higher than either A2 (Spring) or A3 (Summer). Fig. 14b shows a typical result of randomizing[10] the input observation files – i.e. randomly assigning the same observation files to four new lists (A1*-A4*), each containing the same number of files as the original A1 –A4 (see Table 2).

---

[10] For example, 10 observation files (A, B, C, D, … J) are initially grouped in lists A1=[A, B, C, D], A2=[E, F, G], A3=[H, I], A4=[J]. Randomized lists might be: A1*=[B, C, E, I], A2*=[A, F, H], A3*=[D, J], A4*=[G] or A1*=[B, F, G, H], A2*=[A, E, J], A3*=[C, I], A4*=[D] etc.

For each spacecraft season, and for a common low-energy correction for the absorbing galactic hydrogen column ($n_H = 5 \times 10^{20}$ cm$^{-2}$), a power law of the form of eq. 10b was fitted to the spectrum (see Table 3). The spectral slopes differ little from season to season, except for Summer/A3 –where the spectrum is significantly softer. All four photon indices are significantly less than the canonical value of 1.4 and certainly much less than the asymptotic ~2.2-2.4 expected on the basis of Figs. 1a,b. A change of plus or minus 20% to the absorbing column has no influence on the derived photon indices and normalisations. The inclusion of a hydrogen column of course assumes that all contributions to the background are distant from the Earth.

All four seasonal spectra (and their randomised average) do exhibit a "change of slope" or "step feature" at ~2.2 keV, the energy at which the telescope effective area changes discontinuously through the $M_{IV,V}$ edges of the gold mirror coating. The instrument response function does not account for a small excess of counts at an energy of about 2.45 keV. This feature appears more prominent in the A4 spectrum than in its counterpart for spacecraft season A1, as if it were related to the level of the underlying continuum.

The square of the reflectivity – governing the telescope response - of a gold-coated Wolter Type 1 telescope in this energy range is described by Owens et al. (1996) for an angle of grazing incidence of 27 arcminutes. The range of grazing angles for an XMM-Newton telescope is 17.4– 39.7 arcminutes for a point source on-axis. For a uniform diffuse source filling the field-of-view, all nominal grazing angles plus and minus 15 arcminutes are excited simultaneously. As discussed in the Appendix, because X-ray reflectivity decreases with increasing grazing angle even below the critical angle (~1.5 degrees for gold at 2.1 keV), the effective grazing angle for a given mirror shell is decreased for a diffuse source compared to a point source. Measurements made on gold mirror flats for the Astro-H mission by Sugita et al. (2012) indicate that the depth of the Au $M_{IV,V}$ step decreases sharply with decreasing grazing angle in the one degree regime. In the absence of a full ray-trace analysis of an XMM-Newton telescope for a diffuse input, we argue that the match between measured and modelled steps is sufficiently convincing that we may retain the working hypothesis that the seasonal spectra are largely constituted from X-rays originating outside the EPIC pn camera structure, rather than from unrejected soft protons (SPs). We return to the nature of the 2.45 keV peak in Section 4.4.4.

The maximum 2-6 keV pn difference count rate (A4 minus A1, rather than the anticipated Summer minus Winter signal, A3 minus A1) is :

{[0 .421±0.002] - [0.232±0.0015] } = [0.19±0.0025] s$^{-1}$

corresponding to a flux of :

~4.6 x 10$^{-12}$ erg/cm$^2$/s/deg$^2$,

Irrespective of the cause(s) of seasonal variability, the low-state A1/Winter spectrum best represents the true extragalactic CXB, as follows:

$$dF_x/dE = 6.66E^{-0.971} \text{ photons/cm}^2\text{/s/keV/sr} \qquad -(11)$$

## 3.3 Results: EPIC MOS

Figs. 15a,b are the counterparts of Figs. 14a,b for EPIC MOS2. The differences between A1 and the three remaining spacecraft seasons are again apparent, although the counting statistics are poorer than for EPIC pn. The power law fits to the EPIC MOS2 spectra given in Table 3 are broadly consistent, season by season, with those derived for EPIC pn. Fits to the data would be improved by the inclusion in the model of the two strongest Au M-series instrumental lines at 2.122 and 2.203 keV (see Section 3.1.1). There is again a small excess of counts at 2.45 keV, most clearly seen in the randomised (A1*-A4*) MOS2 spectrum of Fig. 15b. The instrumental Chromium line appears, with a seasonally independent intensity, at the expected energy of 5.4 keV. The randomised spectrum departs below 3.5 keV from the "standard" EPIC MOS result for the extragalactic CXB, given in Fig. 3 of De Luca & Molendi (2004). At 2 keV, our normalised count rate is 0.045, compared to 0.07 counts s$^{-1}$ keV$^{-1}$ in that paper.

Fig. 16 shows the spacecraft-seasonal spectra derived from the fully operational EPIC MOS1 CCD array. The derived spectral fits differ significantly in both slope and normalisation from those derived for EPIC MOS2 (see Table 3). The exposure times, however, are a factor of ~2 less. While Winter/A1 once more has the lowest count rate, the three remaining seasons are ordered differently, with A2 and A1 almost identical. Since the observational timelines are very different for the two MOS cameras, this suggests that the variable component of the background does not repeat an annual cycle exactly – as indicated originally in Figs. 5a,b.

Figs.17a,b show the counterparts of Figs.14a,b for EPIC pn split into the two time periods of up to revolution 961 (March 2005) (when MOS1 was fully operational), and afterwards. Though the statistics are poorer, the trend seen in the full-mission data, with the A1 spectrum being low, and the A4 spectrum being high, is still very evident. A very similar situation is seen for MOS2 (not shown), but with poorer still statistics. That the earlier and later behaviours are not identical again points to the variable component of the background not repeating an annual cycle exactly.

## 3.4    Results: Consistency, Significance and North-South Asymmetry

Fig. 18 and Table 2 summarise the 2-6 keV count rates for all three EPIC cameras, after FWC particle background correction, scaled to the X-ray grasp of EPIC pn. The scaling factor which minimises the inter-camera count rate variance is (3.5±0.1), consistent with the approximations to the photon grasp given in Section 2.3.2. The hypothesis that the variable background arises from an external X-ray source which responds to the photon grasp, rather than to the proton grasp (below), is supported. The first basic requirement for consistency between cameras (Section 2.3.2) is met.

Based on the measured seasonal spectra from all three cameras, the null hypothesis that there is *no* "solar axion effect" (i.e that the ordering of spectra – A1 always lowest, A4 always highest - arises by chance) can formally be rejected with a confidence level:

$$1 - (\frac{1}{4})^3 (\frac{1}{3})^3 = 99.94\% \sim 3.1\sigma$$

As alternative measures of significance, Table 3 calculates the difference in power law normalisations for spacecraft seasons A4 and A1, together with the standard error in that difference. This calculation implies that the observed seasonal differences in the X-ray background are significant at the 11σ, 5σ and 4σ levels for EPIC pn, EPIC MOS2 and EPIC MOS1, respectively.

Fig. 19 summarises the significant north-south anisotropy observed by the EPIC cameras. For each camera, the A1 data set was subdivided into files A1(N) and A1(S) and so for the remaining spacecraft seasons, N and S denoting telescope pointing directions north and south of the ecliptic, respectively. In the first half of the mission, restricting the analysis for all three EPIC cameras to revolutions less than 961, the observations indicate a "change-of-sign" – but in Spring, not in Autumn as predicted by the field /orbit model for a single revolution at the beginning of the mission (see Fig. 6). Note that the model does not include the tilt of the geomagnetic dipole; the experimental plane of magnetic symmetry is not actually coincident with the ecliptic plane, at least close to the Earth. Note further that the experimental north-south ratios must include a term, representing the true extragalactic CXB, which is common to both numerator and denominator and which therefore acts to suppress excursions from unity in the measured North-South ratio. Even so, the deviations from isotropy in the seasonally resolved X-ray background appear highly significant. Combining the seasonal data together, all three EPIC cameras show a favouring of the northern hemisphere over the southern by a similar amount (8-20%), at significances from unity of $9.7\sigma$ (pn), $3.9\sigma$ (MOS1) and $4.7\sigma$ (MOS2).

## 4. Discussion

We have presented evidence for a variable component of the much-studied cosmic X-ray background, more than fifty years after the discovery observations by Giacconi et al. (1962). For that new background component to be associated with a dark matter candidate particle which has eluded discovery for most of that time requires that all other possible causes of the variability are carefully considered and conclusively ruled out.

The present study constitutes – at the very least - a new, large-scale study of the CXB (see Table 1) with some significance for the study of AGN evolution and the early Universe. In fact, no independent study of the diffuse X-ray sky with a larger product of grasp and observing time is likely for many years. The best 843 EPIC pn fields alone constitute ~13.8 Ms of sky coverage away from the galactic plane. By comparison, only 85 EPIC MOS 1,2 files corresponding to 34 independent pointing directions at high galactic latitude ($|b| > 28°$) were selected by De Luca & Molendi (2004), a solid angle of only ~5.5 square degrees. The study by Lumb et al. (2002) was based on only eight EPIC MOS fields, while the contemporary Swift XRT study of the CXB by Moretti et al. covered 7 square degrees (2009).

### 4.1 Normalisation and photon index of the randomised spectra

*Do the derived photon indices and normalisations form a credible power law description of the average CXB?*

The recent paper by Moretti et al. (2012) indicates a systematic hardening of the CXB spectrum as fainter and fainter resolved X-ray point source populations are accounted for (see Table 1). A photon index of around unity (Table 3) is consistent with the results of Moretti et al. (2012) for point source fluxes of order $10^{-14}$ ergs cm$^{-2}$ s$^{-1}$ (see Figs. 10a,b). Standard representations of the extragalactic background (Eq. 10a) are based on results from many instruments – non-imaging as well as imaging - whose responses extend to much higher energies than the 6 keV upper energy limit considered here.

We may also look for evidence for a seasonal dependence of the CXB normalisation parameters derived by past surveys (Table 1). For example, the measurements reported by Lumb et al. (2002) were predominantly made in spacecraft season A3, with no observations in spacecraft season A1. Nevalainen et al. (2005) describe a more seasonably balanced programme of observations, with 34% of observing time in A3 and 39% in A1. We then expect (and have confirmed) that the normalisation constant is greater in the first case (11.1) than in the second (7.5).

### 4.2 Difference spectra

*Are the measured difference count rates consistent with calculations based on the axion conversion model outlined in Section 1?*

A proper quantitative model of axion conversion in the real, dynamic, inhomogenous geomagnetic field is not possible, given the unknown nature of the actual differential cross-section for axion-photon conversion. We can, however, estimate to zeroth order the count rates expected in EPIC pn by scaling the curves of Figs. 1a,b by the ratio:

$$[\frac{B_{XMM}^2}{B_{LEO}^2}][\frac{L_{XMM}^2}{L_{LEO}^2}]$$

and using calculated values of the *S*-parameter to represent the average transverse fields in the GECOSAX low Earth orbit (LEO) geometry of Section 1.1 and in the actual XMM-Newton orbit. We ignore the geometric transformation which converts a distant point source to a local isotropic source of X-rays.

Then $35R_E$ and $S/N_{max}$ ~$10^{-6}$ represent the numerator terms and, for the denominator, we calculate $S/N_{max}$ ~0.00014 for a position $0.1R_E$ anti-sunward from the Earth along the GSE x-axis. The scaling factor is then:

875:1

giving maximum scaled count rates well in excess of 1 count/s/keV at 2 keV.

Fig. 20 shows the fit to the EPIC pn "A4 minus A1" difference spectrum of the three-component function $F$ introduced in Section 1.2.1, holding $g_{10}$ constant and varying $g_{11}$. The observed EPIC pn difference signal value of ~0.1 counts/s/keV at 2 keV constrains the value of the axion-electron coupling constant to:

$g_{11}$ ~ 0.22 ± 0.02

if the axion-photon coupling constant $g_{10}$ is constrained to be equal to unity (see Fig. 1b). The resulting estimate of the product of coupling constants:

$g_{ae}g_{a\gamma}$ = 2.2x10$^{-22}$ GeV$^{-1}$

compares with the recently published value 0.8x10$^{-22}$ GeV$^{-1}$ from the CAST team (Barth et al. 2013).

The present analysis also somewhat constrains the axion mass. In a uniform magnetic conversion volume of scale length $L$ =35 $R_E$, the maximum sensitivity is for axion masses

$m_a$ = 2.3x10$^{-6}$ eV

i.e. towards the lower end of the "mass window" described in Section 1.1.1. Given the actual inhomogeneity of the Earth's field, this mass is a lower limit to the actual axion mass. We note that Fermi Large Area Telescope (LAT) GeV gamma ray observations of blazars (Mena & Razzaque 2013) also indicate a low mass (in the range 1 – 3x10$^{-7}$ eV).

The positive detection of solar axions, if confirmed, must have implications not only for our understanding of the true CXB but also for the identification of galactic cold dark matter (CDM). According to Raffelt (2007), an axion mass in the 10 μeV range is sufficient, for a non-thermal dark matter axion population, to account for the entire galactic CDM density.

### 4.3 Alternative Mechanisms

*Is there an alternative mechanism to solar axion-to-X-ray conversion which explains the observed background variation?*

Table 4 lists source mechanisms capable in principle of giving rise to the observed seasonal variation. In terms of observational geometry, only Solar Wind Charge Exchange (SWCX) is expected to have a Summer maximum (Snowden et al. 2004). SWCX X-rays originate in the same local volume of space as the hypothetical axion conversion signal, but are present in the emission lines of highly-ionised C, O, Ne, Mg and possibly Fe (Snowden et al. 2004) below 2 keV, rather than as a broad continuum extending to higher energies. The lack of bright lines or line blends in the observed 2-6 keV spectra argues against an explanation based on the incomplete screening for magnetospheric ion-neutral episodes. As a test, the analyses for EPIC pn and EPIC MOS2 were compared with and without the inclusion of datasets (about 2% of the total) previously believed to show SWCX variability in the energy band containing Oxygen (Carter, Sembay & Read 2010). No significant differences in the 2-6 keV spectra were observed.

Given that the observed seasonal signal certainly appears solar–terrestrial in origin, particular attention does have to be paid to the soft protons which compromise XMM-Newton's general observational programme.

We argue in the Appendix that the single best discriminator between energy deposition by soft protons and the registration of a true, diffuse soft X-ray flux is the ratio of the count rate in EPIC pn to that simultaneously measured in EPIC MOS, because of the very different particle interactions with "front illuminated" and "back illuminated" electrode geometries in the two silicon detectors. The consistency of the pn and MOS count rates with their X-ray grasps (in the ratio ~3.5:1) and not with the ratio of their soft proton grasps – a ratio certainly greater than 5:1 - points to a diffuse photon explanation for the observed variability in the CXB.

The slope of the expected quiescent SP signal spectrum above 2 keV, furthermore, is much less than that actually observed (compare Fig. 20 of the main paper and Appendix Figs. A4 and A5).

### 4.4 Further work

*What more can be done to test the axion hypothesis?*

### 4.4.1 Other X-ray observations

An investigation of "blank sky" background data from the ACIS CCD camera on Chandra (launched in July 1999 into a 133,000 x 16,000 km HEO similar to that of XMM-Newton, but with its apogee in the northern hemisphere) is certainly of great interest. However, the much higher angular resolution of ACIS is accompanied by a significantly smaller peak effective area (~400 versus 800 cm$^2$) and smaller field of view (~256 square arcminutes for ACIS rather than ~700 square arcminutes for EPIC pn), resulting in an X-ray grasp a further factor three below that of EPIC MOS. Chandra studies of the CXB are usually of the "deep pencil beam" form.

The analysis by Hickox & Markevitch (2006) of the Chandra Deep Fields North and South (CDF-N at RA 12h 36min, Dec 62deg 13min; CDF-S at RA 03h 32min Dec -27deg 48min) provided evidence for an unresolved (i.e. truly diffuse) component of the background spectrum which extended to ~5 keV and was significantly anisotropic, in that the count rate was higher in CDF-S than in CDF-N – the opposite of the prediction for XMM-Newton, but just as expected, in the solar axion model, for a spacecraft in an orbit with northern hemisphere apogee. This background component above 1 keV essentially vanished when fainter point source populations detected by the Hubble Space Telescope were added to the background model (Hickox & Markevitch 2007). The Hickox and Markevitch study has been used to set limits on the X-ray narrow line emission from sterile neutrino decay (Abazajian et al. 2007); the EPIC spectra presented above could be used in future to set sensitive limits on the general X-ray background due to such dark decay processes (see also Section 5). Already we can indicate a null detection of solar chameleons, axion like particles (ALPs), whose geomagnetic conversion spectrum is predicted (Brax & Zioutas 2010) to be partly cut off at 2.0 keV (see also Brax, Lindner & Zioutas 2012).

One obvious way to explore the existence of a seasonal component in the CXB is to assemble a sequence of very long X-ray observations of the same field, spaced some months apart. The XMM-Newton observations of the Hubble Deep Field (HDF) originally described by Snowden et al. (2004) may repay further study in this respect.

Observatories in LEO, such as Swift and Suzaku, may not measure a significant seasonal variation in the X-ray background because their magnetic environment is always dominated by the inner dipole region of the Earth's magnetic field. In other words, $B_\perp^2$ is locally large, constant around the spacecraft orbit (at least for low orbital inclinations) and constant with season. The consequence of higher values of the parameter $S$ should be a higher estimate of the *average* X-ray background for an instrument in LEO, compared to that measured by its equivalent in HEO. In practice, any intrinsic sensitivity advantage is probably offset by (i) the lower grasps of Swift and Suzaku compared to XMM-Newton (ii) the variability of the particle background in LEO and the influence of the South Atlantic Anomaly and (iii) Sun and Earth-limb avoidance constraints which preclude lines-of-sight through the bowshock.

### 4.4.2 X-ray polarimetry

New tests of the solar axion hypothesis arise, if, as recently suggested, the conversion X-ray flux is significantly linearly polarised (Payez Cudell & Hutsemekers 2012) - distinguishing it from the true, extragalactic CXB, unpolarised by virtue of its origin in a multiplicity of independent point sources. The recent proposal to the European Space Agency (ESA) for a small satellite (XIPE) may have had the sensitivity to investigate its degree of linear polarisation in an appropriate energy band (Sofitta et al. 2013).

### 4.4.3 Planetary magnetospheres

An independent test of the axion conversion model is available in principle from the X-ray Spectrometer (XRS) (Schlemm et al. 2007) on the MESSENGER spacecraft in (a highly–elliptical) orbit around Mercury. Although the total effective area of the three XRS collimated proportional counters is only ~30 cm$^2$, the aperture is ~100 square degrees, so that the X-ray grasp exceeds that of XMM-Newton EPIC-pn by more than an order of magnitude. The intrinsically weaker field of Mercury, with a dipole moment ~1% that of the Earth, is compensated by the ~6-10 fold increase in axion flux expected from the planet's greater proximity to the Sun. Interpretation of 2-6 keV MESSENGER XRS data may be complicated, however, by the possibility of line emission from calcium and potassium in the exosphere. Furthermore, the XRS's normal mode of operation is pointing to nadir, so the useful observing time may be quite limited. Using the XRS public database[11], a search is underway for any 2-6 keV excess in the stacked spectrum for instrument lines-of-sight intersecting Mercury's bowshock region. The results will be reported separately (Lindsay & Fraser 2014).

### 4.4.4  Axionic line emission

---

[11] NASA Planetary Data System geosciences node http://www.wustl.pds.gov

The appearance in both EPIC pn and EPIC MOS2 data of a narrow feature at an energy of about 2.45 keV recalls the claimed Chandra observation of a line (attributed to the decay of a 5 keV sterile neutrino) in the spheroidal dwarf galaxy Willman 1 (Loewenstein & Kusenko 2010). The claimed line centroid was (2.51±0.11) keV and the flux (3.53±1.95) x $10^{-6}$ photons $cm^{-2}$ $s^{-1}$, both at 90% confidence.

Scaling this flux to XMM-Newton implies an EPIC pn count rate of 0.003 counts/s. In fact, dedicated XMM-Newton observations of Willman 1 (Loewenstein & Kusenko 2012) did not result in a positive line detection. However, the long-term observations of the background presented above *are* consistent with a line of similar count rate at the same centroid energy (see Fig. 21).

Suppose that the agreement between the Willman 1/ Chandra and XMM-Newton/ CXB results is not just coincidence, but represents actual measurements of the same real emission line phenomenon. Then:

(a)     the likelihood of the line being due to an unexpected near-absorption-edge feature of both telescope responses is small; the Chandra High Resolution Mirror Array (HRMA) is coated with iridium, not gold.

(b)     the explanation in the XMM-Newton case cannot be associated with the dark matter halo of a particular distant object; that is, with any particular line-of-sight.

(c)     a line energy of 2.51 keV is close to, but not coincident with, the energies associated with K-shell emission from neutral (2.308 keV), hydrogen like (2.620 keV) and helium-like (2.425, 2.460 keV) sulphur.

Very recently, Redondo (2013) has presented updated calculations of the complex line shapes of the axion production associated with axion-recombination and axion-deexcitation processes involving Ne, Mg, Si, S and Fe ions in the solar core.

If the line feature in question is associated with axion-sulphur interactions, EPIC background spectra should also feature a line associated with silicon, but closer in energy to 1.84 keV than 1.74 keV, and a line associated with iron at 6.70 keV. Fig. 21, showing the folding of the calculated axion spectrum of Fig. 1a plus the line profiles from Redondo 2013 through the EPIC pn instrument response and fitting this to the "A4 minus A1" pn difference spectrum, does indeed indicate the possible presence of emission lines of sulphur at 2.44 and 2.62 keV, and of iron at 6.70 keV. Lines associated with silicon appear less obvious.

The predicted peak axion fluxes at the Earth for narrow-line axion-silicon and axion-sulphur processes are:

$$\sim 1 \times 10^{20} \text{ /m}^2\text{/year/keV}$$

for a $g_{11}$ value of 0.01. Scaling to our continuum result $g_{11}$ = 0.22, then calculating the GECOSAX conversion rate and finally transforming to the XMM-Newton observing geometry gives a detected X-ray line flux of order:

$$\sim 0.1 \text{ counts/s/keV}$$

We would therefore expect to see detected count rates of $\sim 0.001$ $s^{-1}$ in an axionic line feature in EPIC pn spectra if the effective linewidths were of order 10 eV. Calculated line fluxes from the spectral modelling of the EPIC pn difference spectrum, shown in Fig. 21, are 0.0035±0.0005 $s^{-1}$ (2.44 keV), 0.0021±0.0005 (2.62 keV) and 0.0007±0.0004 (6.70 keV), in good agreement with this estimate. Upper limits of 0.0002 $s^{-1}$ and 0.0004 $s^{-1}$ are obtained for the silicon 1.84 and 2.00 keV lines, respectively. A similar situation is seen in both MOS1 and MOS2, with their respective difference spectra showing significant sulphur and iron lines, but weaker silicon features (though the MOS1 1.84 keV line is significant at the 3.6σ level).

A similar count rate can be expected from the 14.4 keV axion line from the M1 nuclear transition in $^{57}$Fe (Redondo 2013; Laming 2013). Such a line would usually be regarded as out-of-band in any EPIC pn analysis because of the fall-off in telescope effective area at high energies, but the detector response is known from pre-launch synchrotron calibration to remain Gaussian up to 15 keV.

Though we do observe a weak feature in the total (A1-A4 combined) EPIC pn spectrum close to 14.4 keV, the poor count statistics at this high energy (where the background is ≈13 times the source signal) prevent a claim for the secure detection of the $^{57}$Fe axion conversion line. Although the laboratory background of the CAST pn CCD camera reported by Kuster et al. (2005) contains the L-series lines of both lead and gold, the EPIC pn spectra lack the strong $L_\alpha$ and $L_\beta$ emission lines of either species, while 14.4 keV lies well below the 17.4 keV energy of instrumental Mo K X-rays (Tenzer et al. 2008), making it unlikely that any candidate line feature, if real, has an origin internal to the detector.

Confirming the 14.4 keV line would be a natural target for the Large Area Detector (LAD) on the recently proposed ESA mission LOFT (Neronov et al. 2013), a collimated instrument with higher detection efficiency above 10 keV, a larger field of view and, above all, enormously greater effective area than XMM-Newton. The grasp of LOFT/LAD is almost three orders of magnitude higher than that of EPIC pn above 10 keV.

## 5. Conclusions

Conlon and Marsh (2013) postulate that the decay of string theory moduli will give rise to a Cosmic Axion Background (CAB), whose interaction with intergalactic magnetic fields gives rise to an observable fraction of the 0.1-1 keV CXB. The maximum signal is predicted at an X-ray energy of 0.2 keV. This work has in turn given rise to a paper (Fairbairn 2014) which further explores the connection between CAB and CXB, asking:

*Could these photons coming from axions explain the cosmic X-ray background?*

This paper appeared, therefore, to be a timely contribution to the debate on the nature of dark matter, even before the very recent studies indicating the presence of a 3.55 keV sterile neutrino decay line in the stacked spectra of clusters of galaxies (Bulbul et al. 2014; Boyarsky et al. 2014). An axion-related line complex arises in this energy range from potassium or argon in the solar core, but is not obvious in Figs. 14, 15 and 16. Nevertheless, the detection of even two simultaneous dark matter lines poses severe difficulties for a dark matter model based on the decay of a solitary candidate particle.

On the basis of our results from XMM-Newton, it appears plausible that axions – dark matter particle candidates - are indeed produced in the core of the Sun and do indeed convert to soft X-rays in the magnetic field of the Earth, giving rise to a significant, seasonally-variable component of the 2-6 keV CXB. The confirmation of narrow axionic line features associated with silicon, sulphur and iron, in addition to a continuum exhibiting seasonal variation and north-south anisotropy, would raise the bar very high against competing explanations.

## 6. Acknowledgements

provided a critical reading of an early draft manuscript. We are very grateful to Alvaro de Rujula for very fruitful discussions, and we also thank the referee for very useful comments, which have improved the paper. The work of AMR, SS and JAC on the calibration of XMM-Newton was supported by the UK Space Agency.

**Table 1**
Spectral form of the Cosmic X-ray Background from EPIC MOS, EPIC pn, Swift XRT and Chandra ACIS data sets. Distribution of observation times by spacecraft season [A1, A2, A3, A4] (as percentages) are given in rightmost column.

| Reference | Excluded Point Source Flux Limit (erg cm$^{-2}$ s$^{-1}$) | Best-fit Photon Index | Normalisation (ph cm$^{-2}$ s$^{-1}$ sr$^{-1}$ keV$^{-1}$ at 1 keV) | Notes |
|---|---|---|---|---|
| Lumb et al. (2002) | (a) No exclusion (b) 1-2 x 10$^{-14}$ | (a) 1.45 (b) 1.42 | (a) - (b) 11.1 | 1.2 square degrees sky coverage [0, 15, 73, 12] |
| De Luca & Molendi (2004) | Only bright target source excluded. | 1.41±0.06 | 11.6 | 5.5 square degrees at high galactic latitudes. |
| Carter & Read (2007) | ~10$^{-14}$ | 1.37±0.15 1.45±0.15 1.42±0.07 1.50±0.19 | | Independent estimates from separate halves of MOS1 and MOS2 [15, 24, 33, 27] [21, 06, 45, 28] |
| Moretti et al. (2009) | Excluding only GRB afterglow target | 1.47±0.07 | 12.2 | 126 GRBs; 7 square degrees. |
| Snowden, Collier & Kuntz (2004) | | 1.46 | | Discovery of SWCX in sequence of Hubble Deep Field (North), June 1-2 (2001) [0, 60, 40, 0] |
| Nevalainen et al. (2005) | | | 7.5 (derived) | Concludes quiescent SP level is non-zero [39, 12, 34, 15] |
| Hickox & Markewitz (2007) | | | | Chandra Deep Fields North and South |
| Abazajian et al. (2007) | | | | Sterile neutrino line emission search using Chandra Deep Fields |
| Soltan (2007) | As Lumb (2002) | 1.42 | 9.0 | Spatial variation of background components measured using auto correlation function |
| Moretti et al. (2012) | No exclusion 5 x 10$^{-15}$ 3 x 10$^{-16}$ 1x 10$^{-17}$ | 1.20±0.05 0.84±0.1 0.7±0.3 0.1±0.7 | . . . 0.25 | Combined Swift XRT / Chandra Deep Field –South analysis |

**Table 2**
EPIC pn and MOS file statistics per spacecraft season. Total counts are in the 2-6 keV band before FWC cosmic ray background correction. Selected files have flux-in to flux-out ratios of $0.95 < R < 1.3$, and for pn and MOS2, all revolutions are used, while for MOS1, up to revolution 961 only is used. The count rate ratios computed in the rightmost two columns should be compared with the ratio of X-ray grasps $G_{pn}/G_{MOS}$ (see Section 2.3.2).

| Camera | S/C Season | Number of Best Files | Exposure time (Ms) | Total Counts (x$10^6$) | Average background rate (s$^{-1}$) | Average CXB rate (s$^{-1}$) | Background ratio pn/MOS | CXB ratio pn/MOS |
|---|---|---|---|---|---|---|---|---|
| pn | A1 | 308 | 5.3 | 5.34 | 1.01 | 0.23 | | |
| | A2 | 142 | 2.47 | 2.75 | 1.11 | 0.31 | | |
| | A3 | 248 | 3.62 | 3.88 | 1.07 | 0.30 | | |
| | A4 | 148 | 2.57 | 3.28 | 1.30 | 0.42 | | |
| MOS2 | A1 | 336 | 7.17 | 2.59 | 0.36 | 0.08 | 2.80 | 3.30 |
| | A2 | 174 | 3.66 | 1.39 | 0.38 | 0.01 | 2.95 | 3.26 |
| | A3 | 287 | 5.31 | 2.05 | 0.39 | 0.10 | 2.78 | 2.88 |
| | A4 | 194 | 4.24 | 1.75 | 0.41 | 0.11 | 3.13 | 3.86 |
| MOS1 | A1 | 130 | 2.85 | 0.83 | 0.29 | 0.07 | 3.46 | 3.33 |
| | A2 | 79 | 1.38 | 0.42 | 0.30 | 0.08 | 3.30 | 4.02 |
| | A3 | 122 | 2.13 | 0.63 | 0.30 | 0.07 | 3.64 | 3.98 |
| | A4 | 64 | 1.32 | 0.44 | 0.33 | 0.10 | 3.92 | 4.21 |

**Table 3**
Best fit photon indices Γ and normalisations $N_0$ (cm$^{-2}$ s$^{-1}$ keV$^{-1}$ sr$^{-1}$) for the spectral fits to the EPIC pn, MOS2 and MOS1 stacked spectra, per spacecraft season. The rightmost column gives, for each of the EPIC cameras, a calculation of the significance of the seasonal variation, based on the difference between the normalisation terms for A4 and A1.

| EPIC Camera | Season | $N_0$ | Γ | |
|---|---|---|---|---|
| pn | A1 | 6.66+0.23 -0.32 | 0.97±0.03 | Difference 12.09-6.66 = 5.43 Uncertainty $(0.32^2 + 0.36^2)^{1/2}$ = 0.48 Significance 5.43/0.48 = 11.4σ |
| | A2 | 9.08+0.25 -0.35 | 0.98±0.03 | |
| | A3 | 9.60+0.19 -0.43 | 1.06±0.03 | |
| | A4 | 12.09+0.26 -0.36 | 0.97±0.03 | |
| MOS2 | A1 | 7.05+0.45 -0.29 | 0.91±0.05 | Difference 10.50 – 7.05 = 3.45 Uncertainty $(0.45^2 + 0.52^2)^{1/2}$ = 0.69 Significance 3.45/0.69 = 5.0σ |
| | A2 | 8.79+0.31 -0.46 | 0.90±0.04 | |
| | A3 | 10.32+0.50 -0.32 | 0.98±0.03 | |
| | A4 | 10.50+0.52 -0.34 | 0.94±0.03 | |
| MOS1 | A1 | 5.43+0.34 -0.40 | 0.72±0.06 | Difference 8.26– 5.43 = 2.83 Uncertainty $(0.40^2 + 0.57^2)^{1/2}$ = 0.70 Significance 2.83/0.70 = 4.0σ |
| | A2 | 6.68+0.57 -0.36 | 0.80±0.05 | |
| | A3 | 7.54+0.61 -0.38 | 0.93±0.05 | |
| | A4 | 8.26+0.57 -0.37 | 0.77± 0.04 | |

**Table 4**
Alternative possible contributions to a seasonally variable 2-6 keV X-ray background.

| Mechanism | Comments | Reference |
|---|---|---|
| Solar Wind Charge Exchange (SWCX) | Observational geometries for SWCX and axion conversion X-rays very similar. All known XMM-Newton SWCX data sets removed from present analysis. SWCX spectra line-dominated below 2 keV, incompatible with observed line-free, broad continuum. | Snowden Collier Kuntz (2004) |
| Galactic background, local bubble | Significant only below ~2 keV. Galactic plane excluded. No correlation of season A1 or A4 observations with galactic poles. | Galeazzi et al. (2011) |
| Residual soft proton (SP) flux | Systematic removal of soft proton fluxes from EPIC pn and EPIC MOS data using established methods. | Appendix |
| keV electron flux | Times are known when XMM orbit lies within trapped belts or magnetotail plasma sheet. Narrow peaks in electron flux ~10-20 orbits wide associated with plasma sheet crossings, rather than broad seasonal variations. | Rosenqvist et al. (2002) |
| Earth albedo | Earth X-ray albedo spectrum peak luminosity at ~40 keV; dayside argon K-shell fluorescence line expected at 2.96 keV, not observed in present work- nor in Suzaku dark Earth observation | Churasov et al. (2008) |
| Compton-Getting effect / X-ray dipole | Anisotropy of the CXB due to motion relative to the distant Universe convolved with pointing constraints could mimic seasonal variability. Measured spatial variation only at ~2% level. | Revnivtsev et al. (2008) |

.

# Figure Captions

**1.(a)** Solar axion conversion X-ray spectra in the GECOSAX observing geometry. Calculated from Eqs. 1-5 with $L= 600$ km, $g_{10} = g_{11} = 1$, $A = 800$ cm$^2$ and the values for $k_1$ and $k_2$ given in the text. Full curves labelled with axion generation mechanisms (Primakoff, Compton and Bremsstrahlung); the fourth curve is the sum of the Compton and Bremsstrahlung signals. A power law approximation to the sum signal has an index 2.5 at 6 keV, steeper than the 1.4 of the usual CXB model spectrum.
**(b)** Total conversion X-ray spectra for $g_{10} =1$ and (top to bottom) $g_{11} = 1, 0.5, 0.2$ and $0.1$.

**2.** Schematics in the GSE noon-midnight plane of the orthogonal GECOSAX scenario, of the XMM-Newton orbit (elongated ellipse) and of the Earth's magnetic field, represented by field lines with McIlwain L-values 5, 10 and 20 and by a parabolic approximation to the magnetopause. The broken circle indicates the 40,000 km minimum altitude for XMM-Newton observations. The shaded triangle indicates the allowed pointing directions for XMM-Newton (approximately +/-20 degrees from perpendicular to the Sun-Earth line); the triangle can 'rotate' about the X-axis (the Sun-Earth line). X-ray conversion photons can enter the field-of-view of XMM-Newton via various mechanisms, discussed in the text, shown schematically as (a) elastic scattering, (b) non-co-linear conversion plus axion-to-photon back-conversion, and (c) non-co-linear conversion. **(top)** Winter configuration, 1$^{st}$ January 2000. **(bottom)** Summer configuration, 1$^{st}$ July 2000.

**3.(a)** Projection of the XMM-Newton orbit onto the ecliptic plane, at three-monthly intervals at beginning of mission. The central circle represents the minimum altitude for observations. The apogee of the orbit on January 1$^{st}$ 2001 is indicated by the radial line. The spacecraft "Winter solstice", when the apogee of the orbit lies on the Sun-Earth line, falls in mid-January 2001. **(b)** As Fig. 3a, for mid-mission period 2006-7. Precession of the orbit means that the spacecraft Winter solstice now occurs in late November. The Sun is far to the right at X=23450 $R_E$, Y=0 $R_E$.

**4**. Logarithmic contour plot of the square of the transverse B-field of the Earth in the GSE noon-midnight plane.

**5.(a)** Signal *S* calculated at monthly intervals beginning July 1$^{st}$, 2000, near the start of the mission. Open and filled circles indicate signal values before and after correction for the variation of solar axion flux due to the eccentricity of the Earth's orbit. A broad Winter/Spring minimum is predicted for the axion conversion X-ray signal.
**(b)** As Fig. 5a, but beginning in mid-mission, 1$^{st}$ July 2006 (~Revolution 1000). Squares and right hand scale – RGS 2000-8 average high energy background count rates for the quietest phases of the XMM-Newton orbit, 18-24 hr (open symbols) and 24-30 hr (filled symbols) after perigee (Rodriguez-Pascual & Gonzalez-Riestra 2008). The axion signal is predicted to have a strong Autumn/A4 peak. The calculated minimum in *S* is correlated with the January/February minimum in the RGS data and corresponds to the season when XMM-Newton is physically inside the Earth's magnetosphere for much of the time (see Appendix Fig. 1).

**6.** Signal *S* (left hand scale) calculated at monthly intervals, beginning on July 1$^{st}$ 2000 for northward (x symbols) and southward (+ symbols) pointing directions. The horizontal bars, full curve and right hand scale indicate the seasonal north-to-south signal ratios. The largest asymmetry is in Spring/A2; the north-south dependency reverses in spacecraft Autumn.

**7.** Extremes of XMM-Newton orbital path versus revolution number, from the start of the mission (revolution 0; December 1999) to revolution 2360 (October 2012). Small scale curves – perigee coordinates, in units of 100,000 km. Large scale curves – apogee coordinates. Red symbols – $X_s$ coordinate; blue – $Y_s$ and green – $Z_s$. The $Y_s$ and $Z_s$ coordinates of the apogee exhibit an annual variation overlaid with a long-period (~20 year) modulation. The spacecraft elongation from the Earth measured by the $Y_s$ coordinate of the apogee is larger at the start of mission, implying better "visibility" of the sunward axion conversion volume. The coloured bars at the base of the figure indicate the spacecraft seasons. Here, as in subsequent figures: A1 – black; A2 – red; A3 – green; A4 – blue.

**8.** Filter wheel closed (FWC) scaling factors versus revolution number. **(a)** EPIC pn. **(b)** EPIC MOS2. For both cameras, the datasets for the individual spacecraft seasons (black, red, green and blue circles) follow a common curve – which peaks around the 2009-2010 solar minimum. Revolution 200 was on 10$^{th}$ January 2001 and revolution 2000 was on 9$^{th}$ November 2010.

**9.** Exposure time histograms. **(a)** EPIC pn **(b)** EPIC MOS2.

**10**. Flux histograms for removed point sources. **(a)** EPIC pn. **(b)** EPIC MOS2.

**11.** Changes in relative seasonal distribution of "best" data files with mission elapsed time. **(a)** EPIC pn**. (b)** EPIC MOS2.

**12.** Sky distribution of observations in Galactic projection. The Galactic plane (excluded) is shown by the horizontal dashed line. An ecliptic coordinate grid is plotted, with the North Ecliptic pole (NEP) and South Ecliptic pole (SEP), and the ecliptic plane (line running bottom-left to top-right) marked (along which the Sun is positioned). **(a)** EPIC pn**. (b)** EPIC MOS2.

**13**. Distributions of data files by *R* value. **(a)** EPIC pn. **(b)** EPIC MOS2 The range 0.95-1.3 is regarded as conservative in terms of soft proton flare rejection.

**14. (a)** EPIC pn X-ray background spectra for each of the four spacecraft seasons, with A1/Winter (black symbols) clearly differentiated from A4/Autumn (blue symbols), A2/Spring (red) and A3/Summer (green). Full curve – power law fit to low state A1/Winter data (see Table 3).
**(b)** EPIC pn randomised X-ray background spectra, all following a single common curve. Here the same input observation files as in (a) are randomly assigned to four new lists (A1*-A4*), each containing the same number of files as the original A1–A4 in (a). These lists are then used in the same way as for (a) to generate the randomized spectra shown in (b). A typical example of the randomised spectra is shown.

**15. (a,b)** As Figs. 14(a,b) for EPIC MOS2. Note the seasonal independence of the chromium instrumental line at 5.4 keV, contrasting with the behaviour of the narrow excess feature at ~2.45keV, whose significance increases with the local continuum count rate.

**16 (a).** As Fig. 14(a) for EPIC MOS1, and limited to the first half of the mission, up to revolution 961.

**17. (a,b)** As Figs. 14(a,b) for EPIC pn split into the time periods (a) pre- and (b) post-revolution 961.

**18.** Scaled background count rates in the 2-6 keV energy band versus spacecraft season for all three EPIC cameras ; scaling is to the photon grasp *G* of EPIC pn. A scaling factor (3.5:1) applied to the EPIC MOS1 and EPIC MOS2 count rates minimises the overall variance between cameras/season combinations. The error bars are smaller than the individual symbols.

**19.** North/South 2-6 keV count rate ratios versus spacecraft season, for revolution numbers less than 961. Open symbols – red circles, EPIC pn; green diamonds, EPIC MOS1; blue squares, EPIC MOS2. The dashed horizontal black line at y=1.0 indicates a perfectly isotropic X-ray background. The coloured dashed lines and filled symbols indicate the North/South ratios for the EPIC pn (red), MOS1 (green) and MOS2 (blue) data, after combining the seasonal data together.

**20**. Comparison of EPIC pn difference spectrum (individual symbols) with axion conversion spectra calculated for stated values (0.2, 0.22, 0.24) of the axion-electron coupling constant.

**21**. EPIC pn difference spectrum (A4 – A1) in the 1.85-7.5 keV band, fit with a model comprising the Bremsstrahlung (blue) and Compton (red) components of the expected Solar axion conversion spectrum of Fig. 1a, plus added line features centred on 1.84, 2.00 (silicon), 2.44, 2.62 (sulphur) and 6.70 keV (iron). The sulphur and iron lines are visible at the figure base.

**Figs. 1a (top) and 1b.**

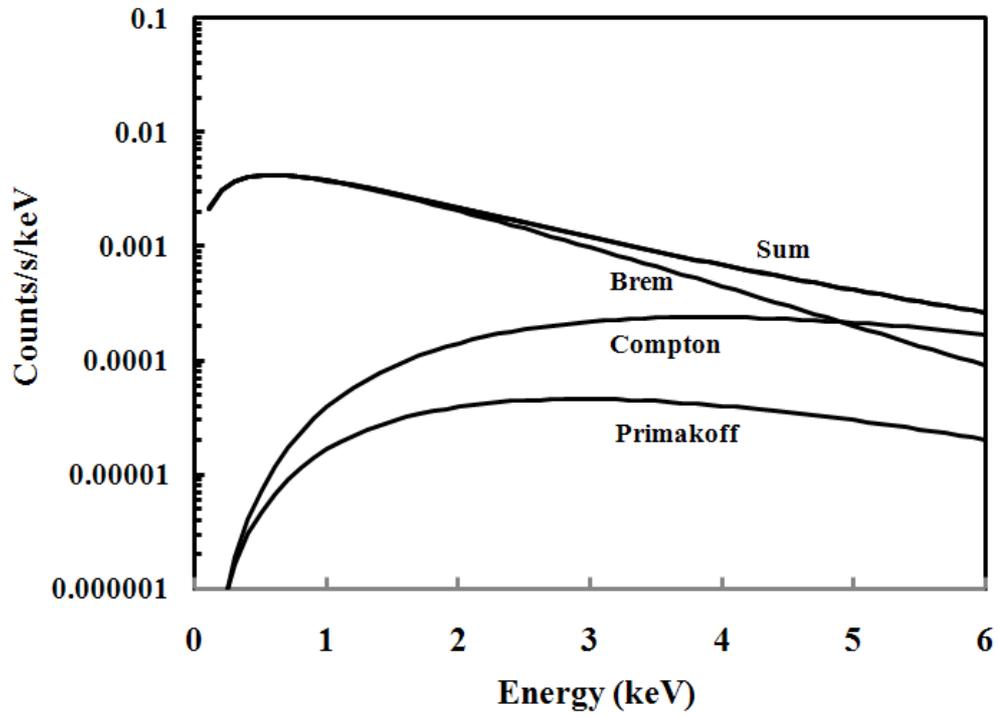

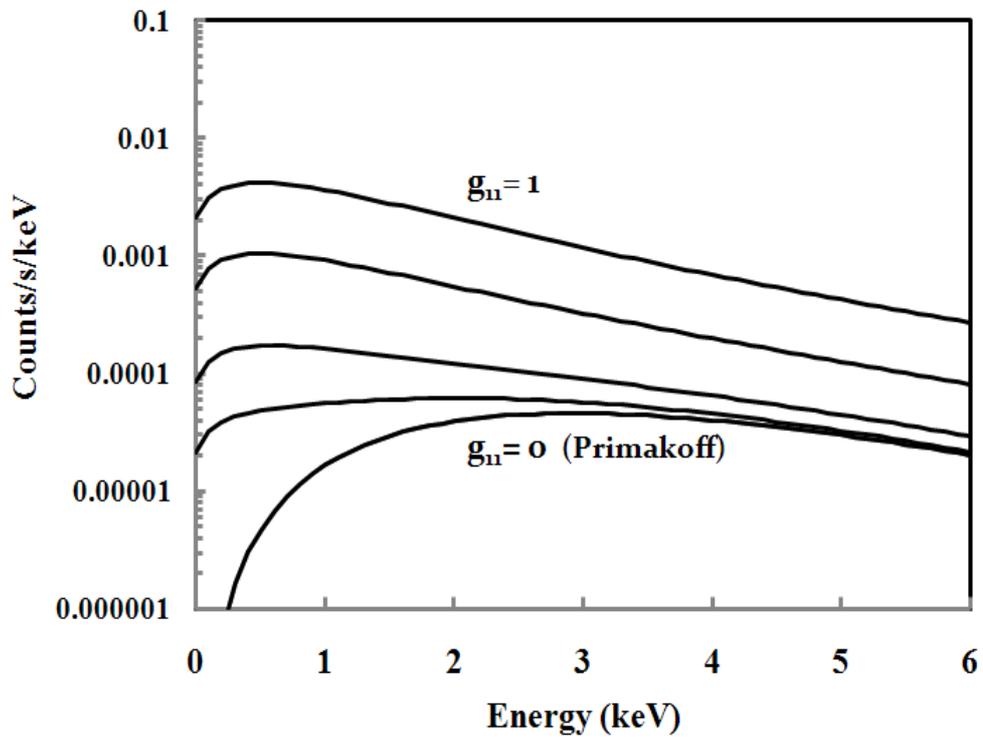

**Figs. 2a (top) and 2b.**

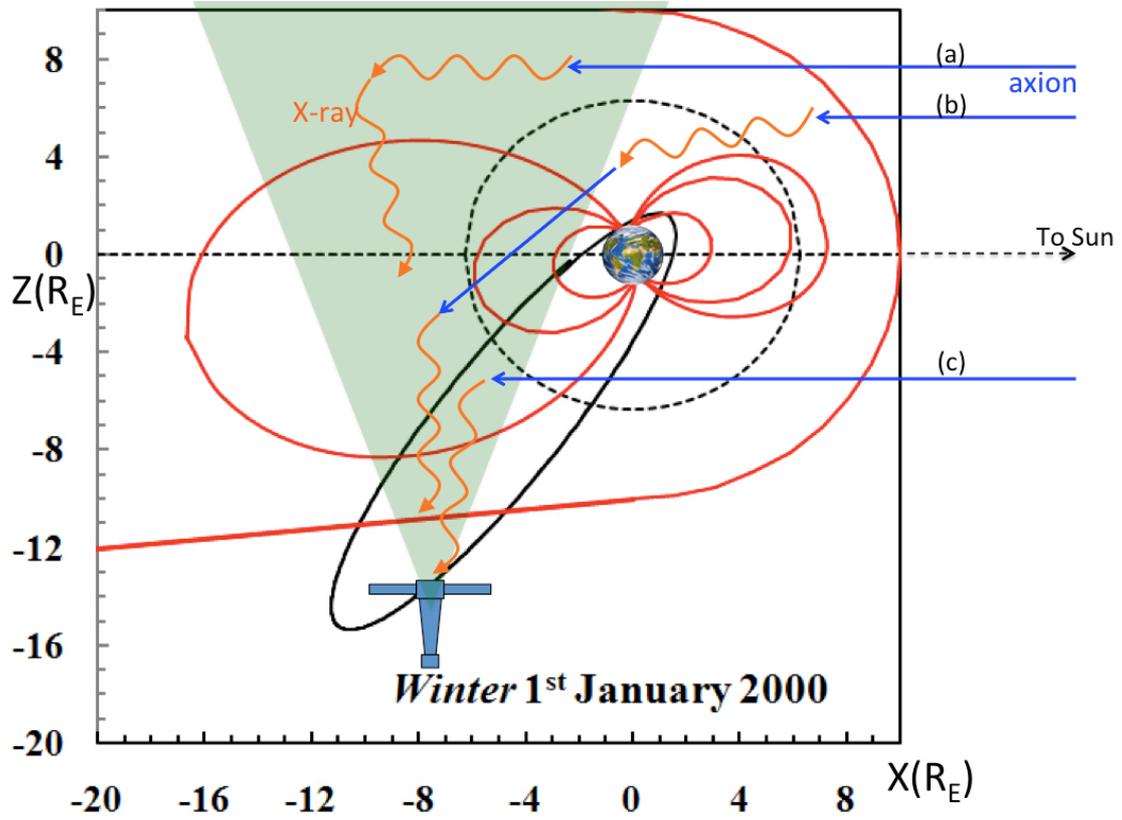

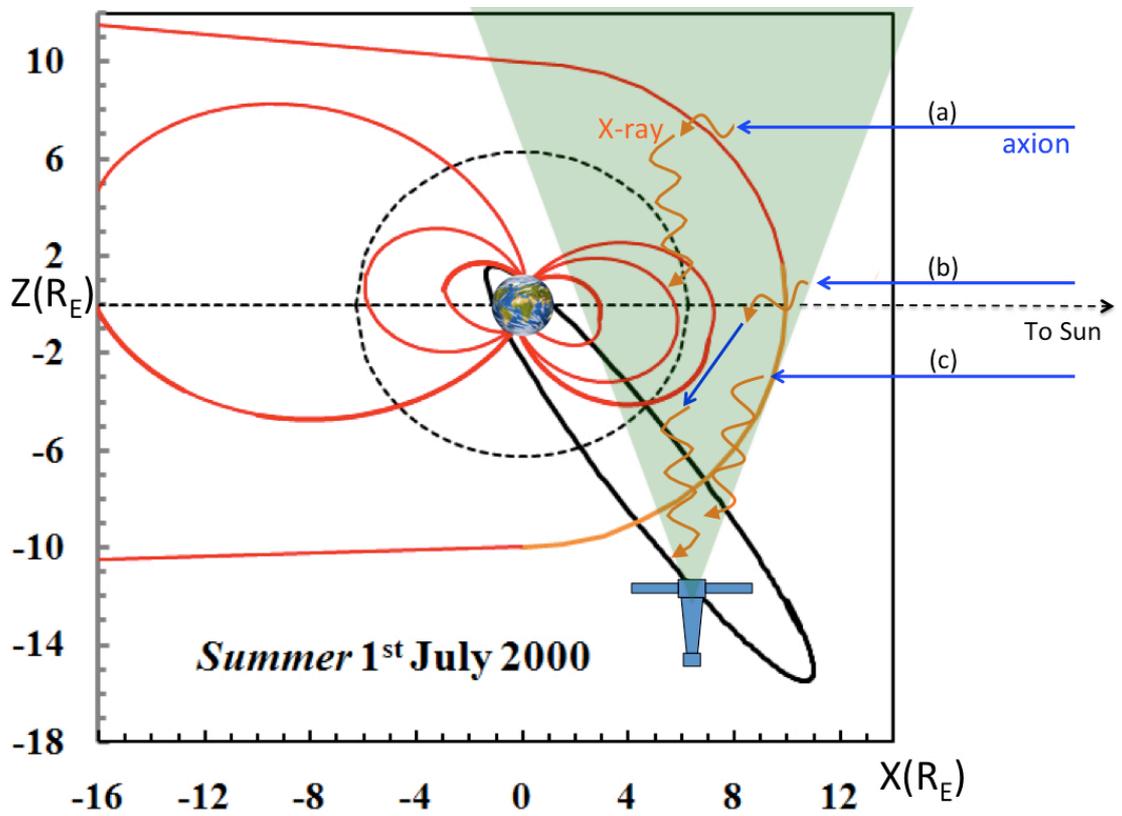

**Figs. 3a (top) and 3b.**

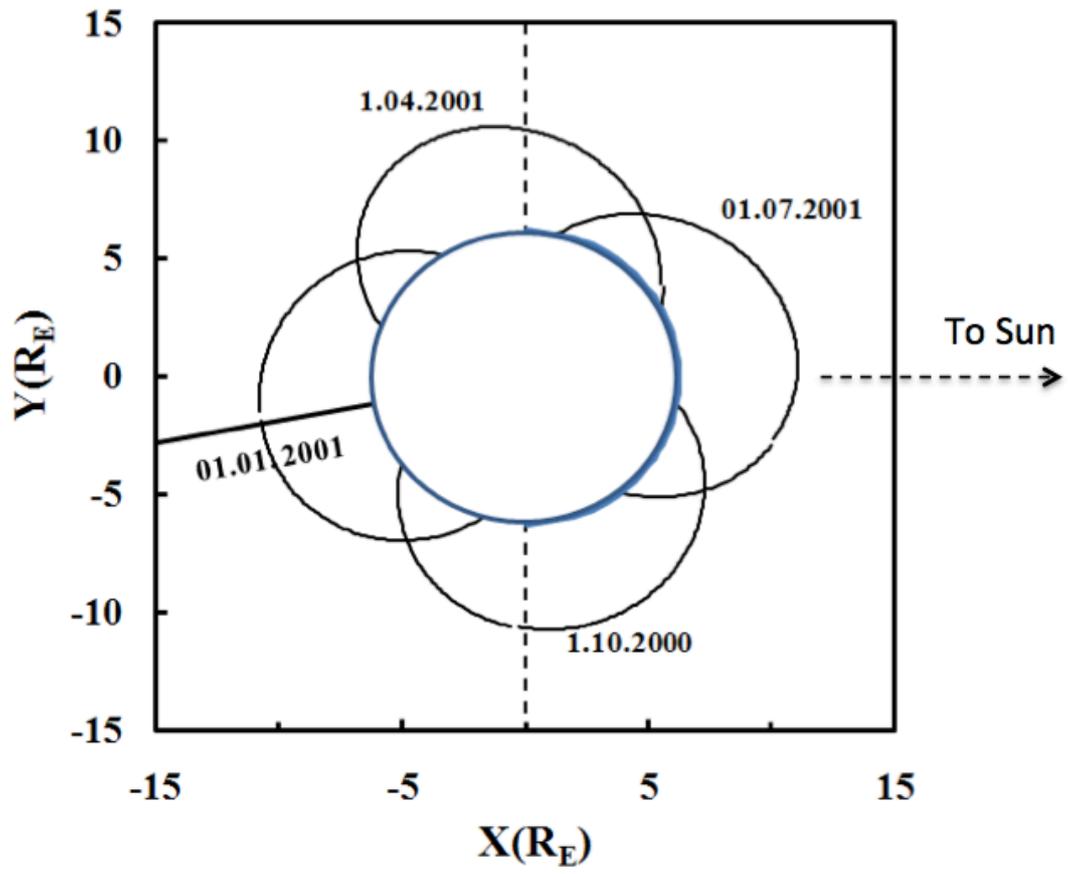

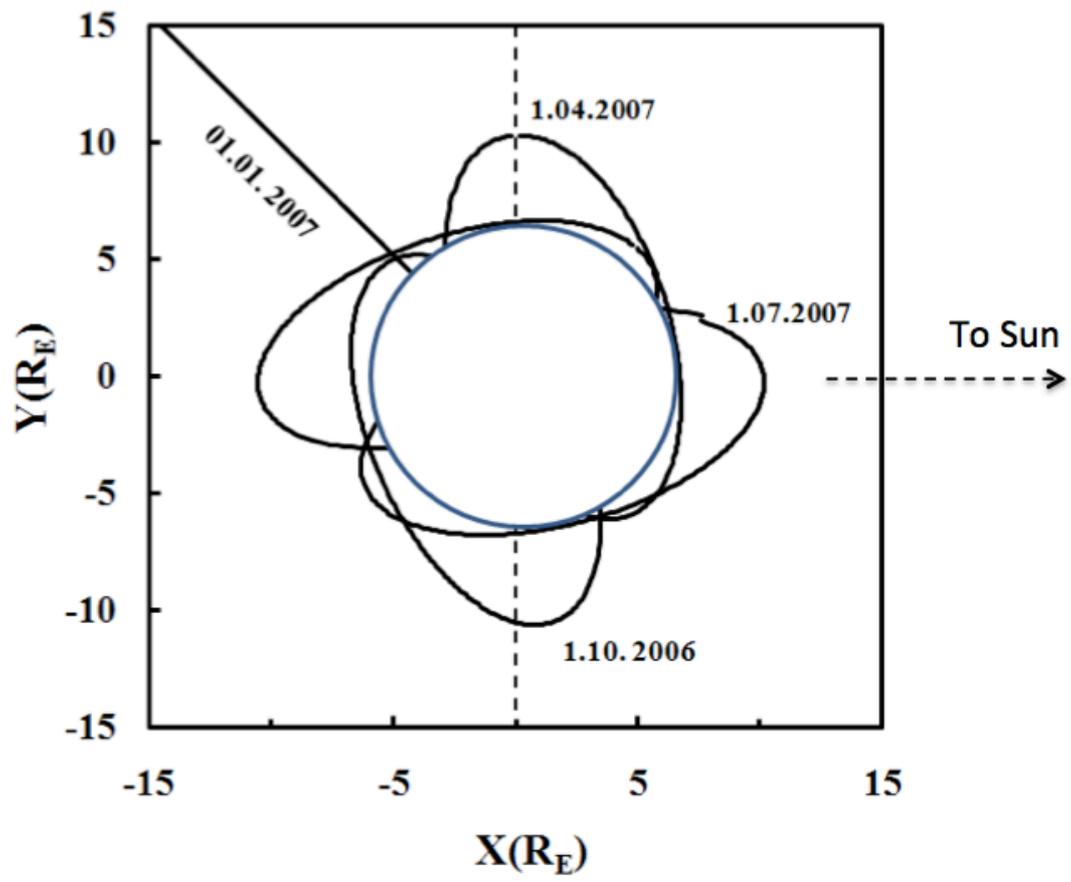

**Fig. 4.**

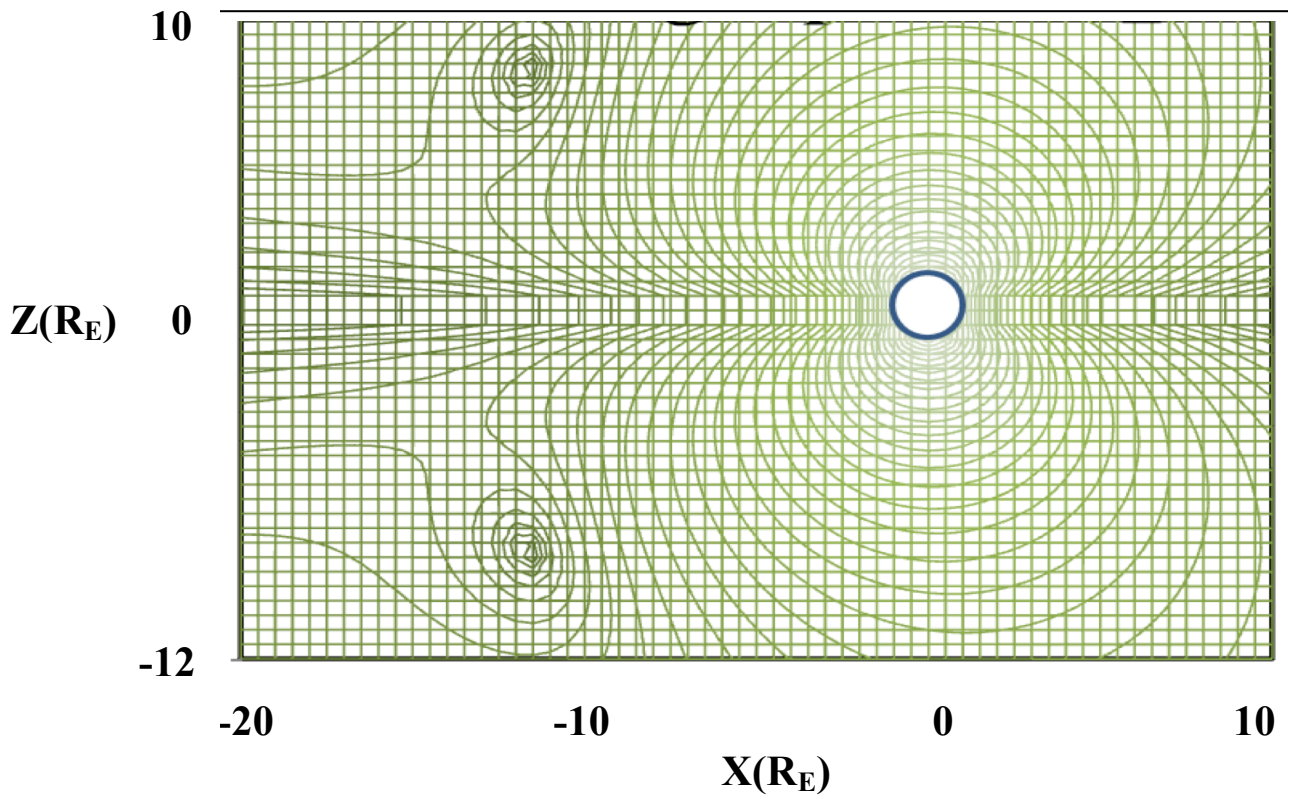

**Figs. 5a (top) and 5b..**

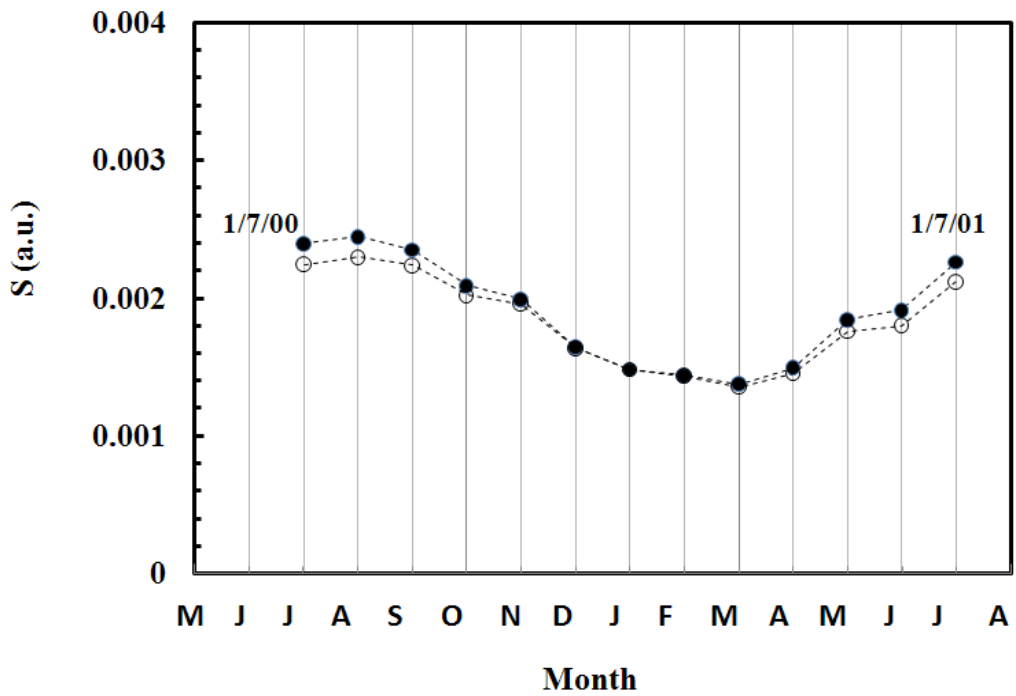

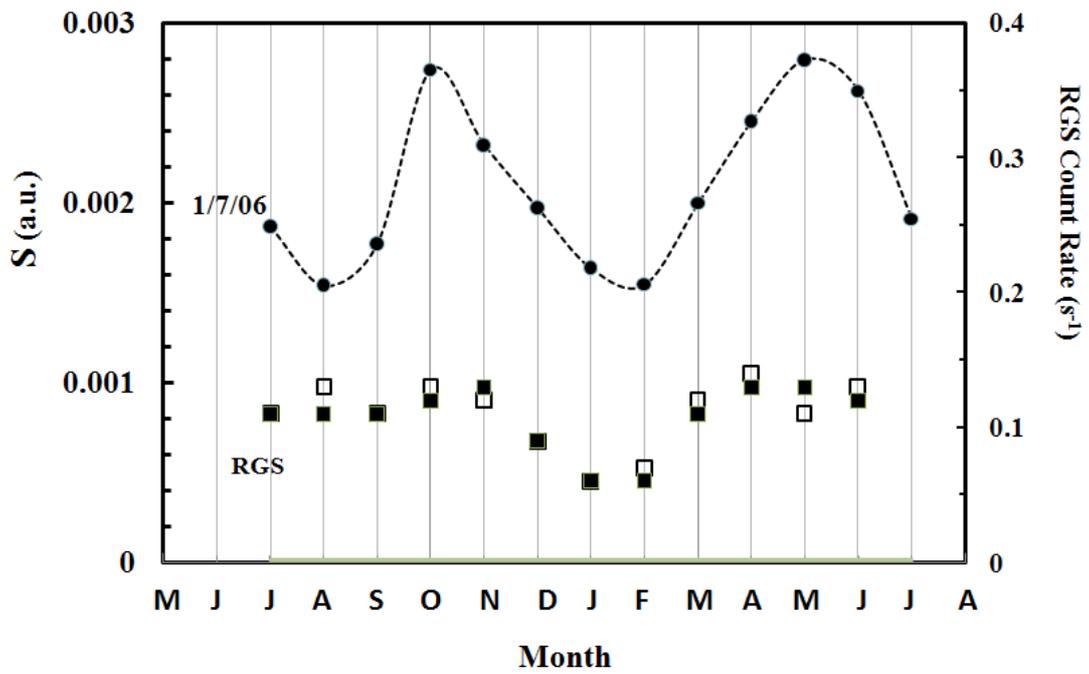

**Fig. 6.**

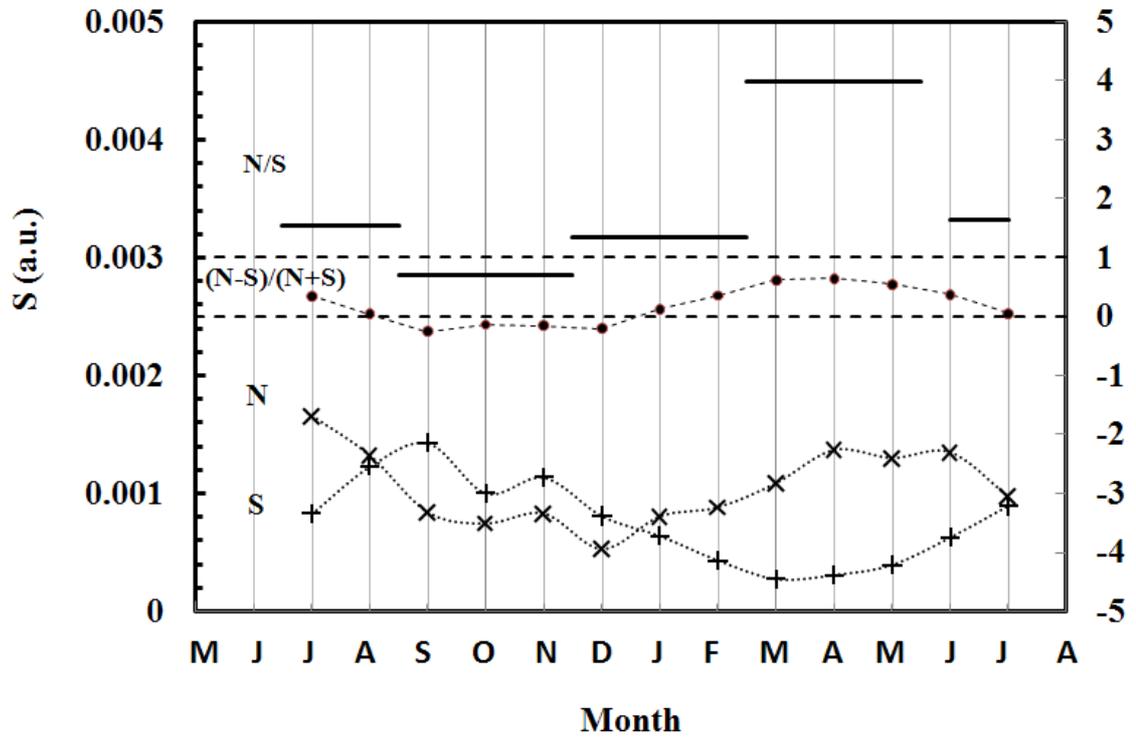

**Fig. 7**.

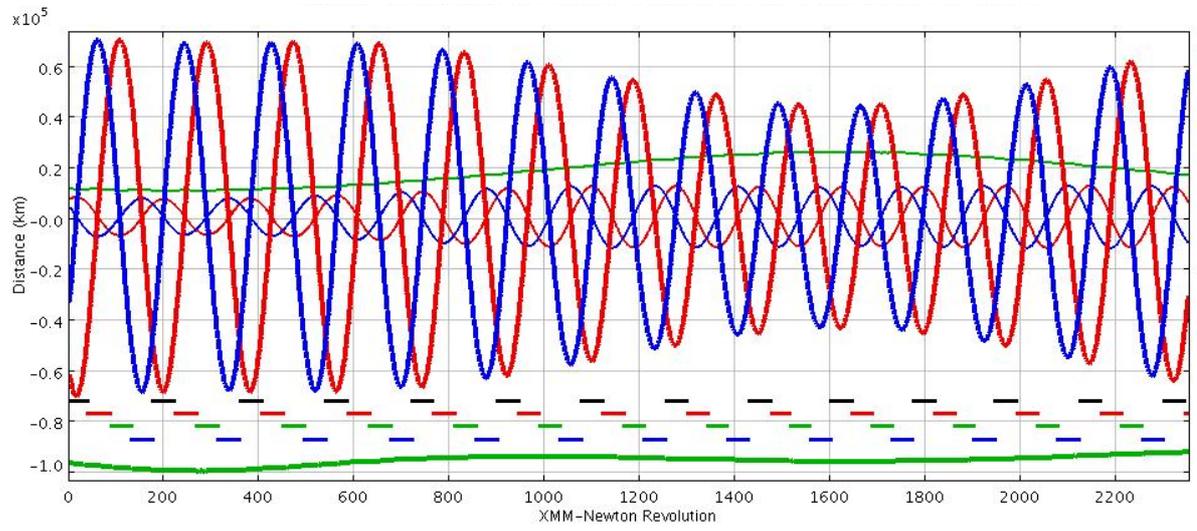

**Fig. 8a.**

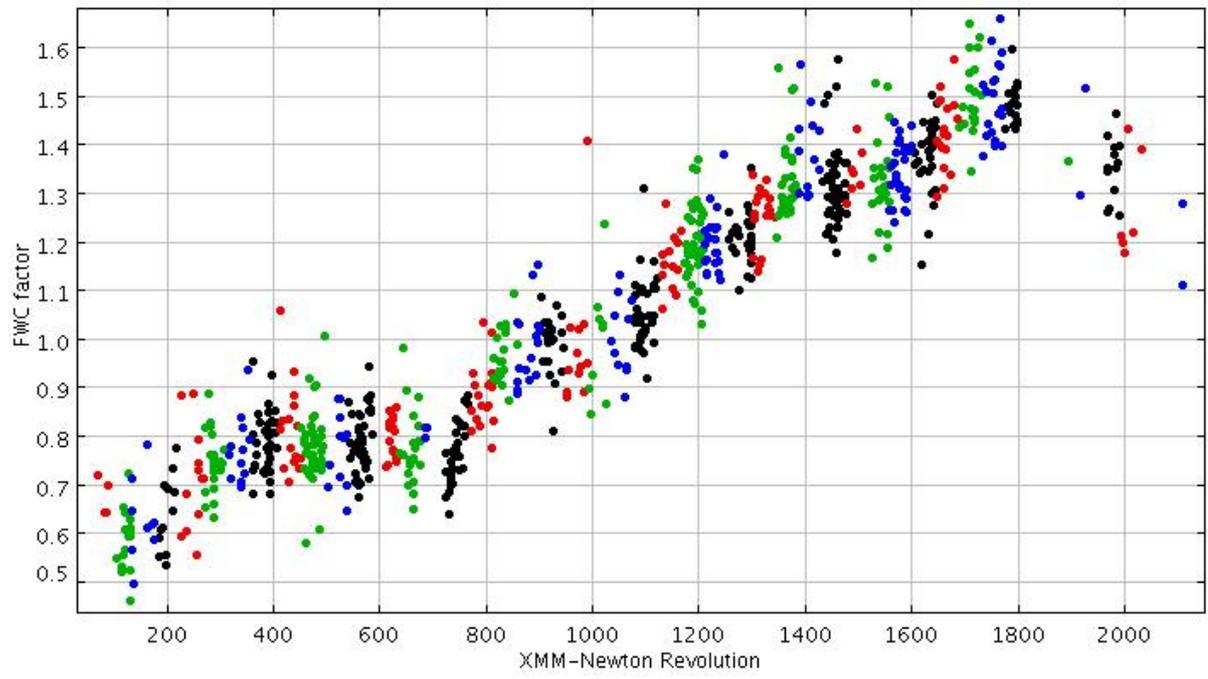

**Fig. 8b**.

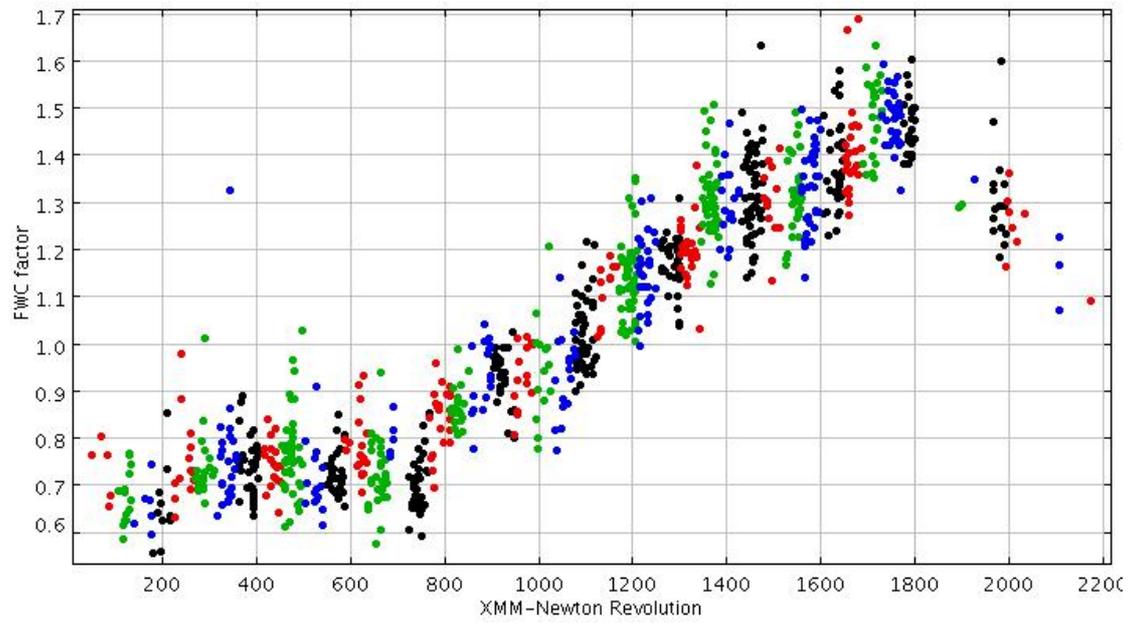

**Figs. 9a (top) and 9b.**

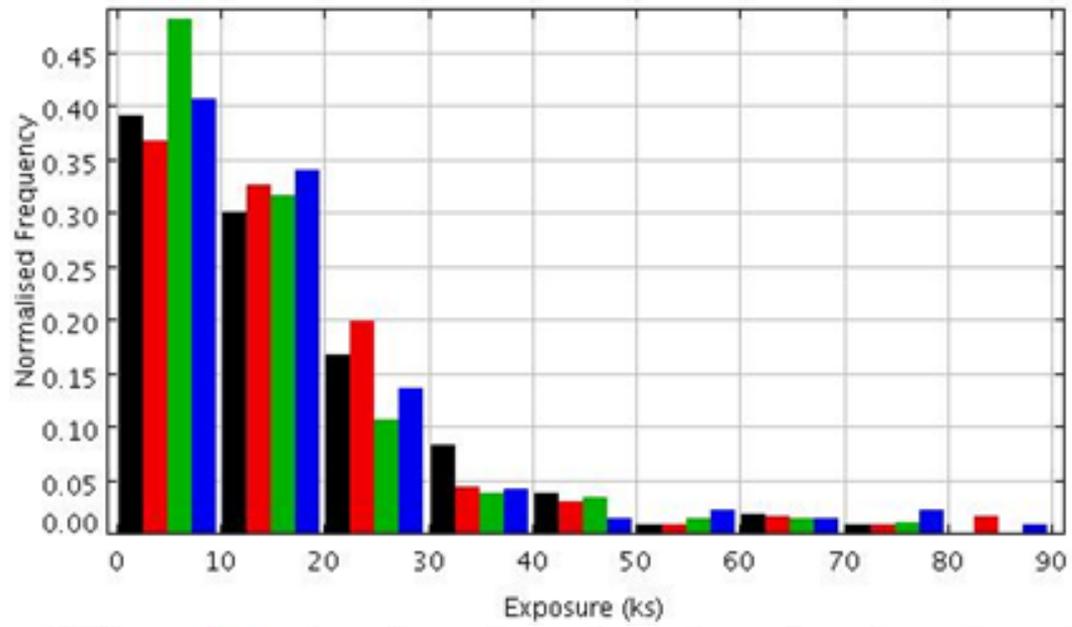
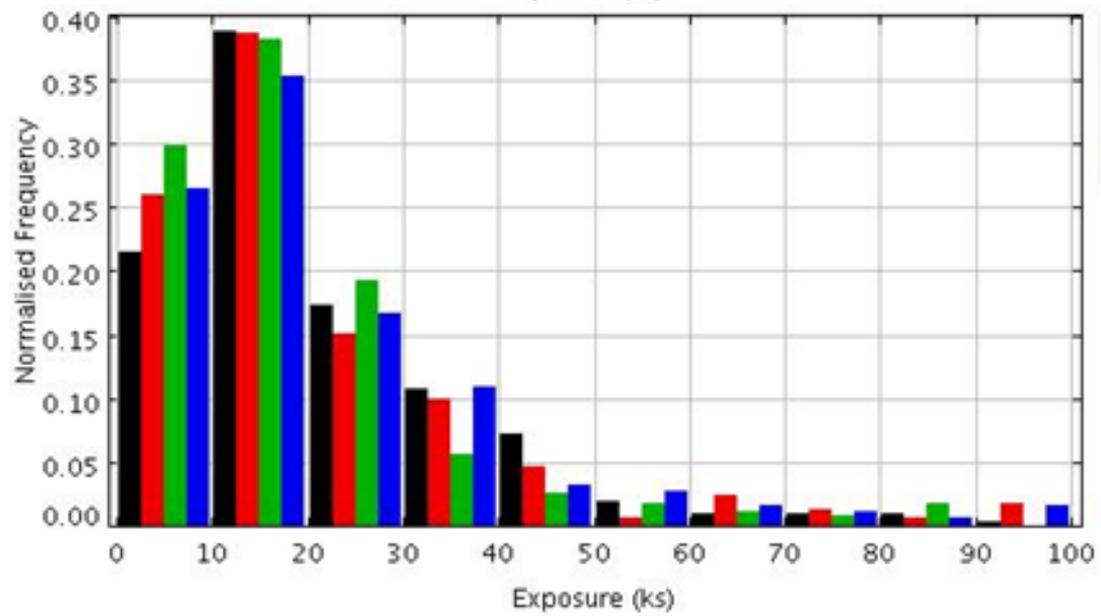

**Fig. 10a (top) and 10b.**

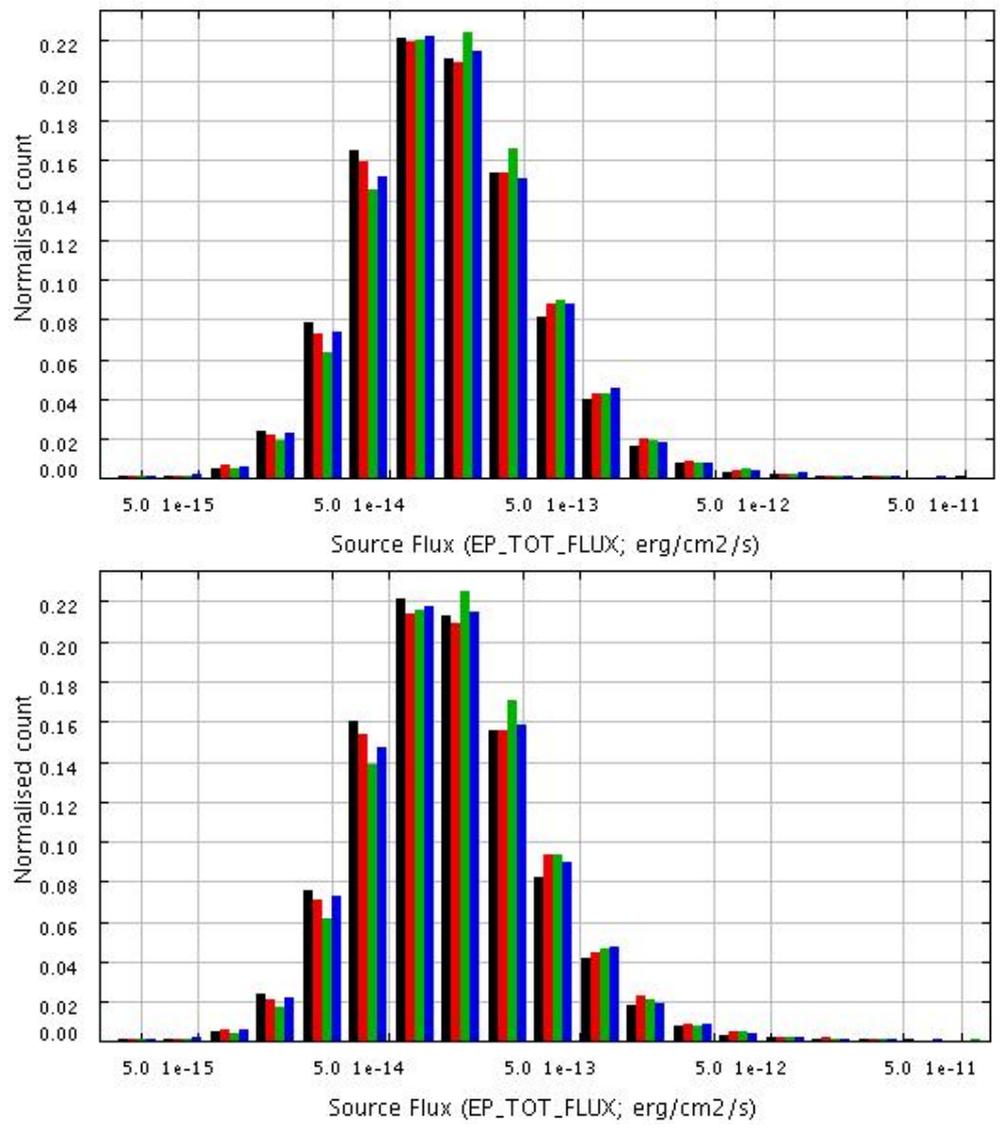

**Fig. 11a (top) and 11b.**

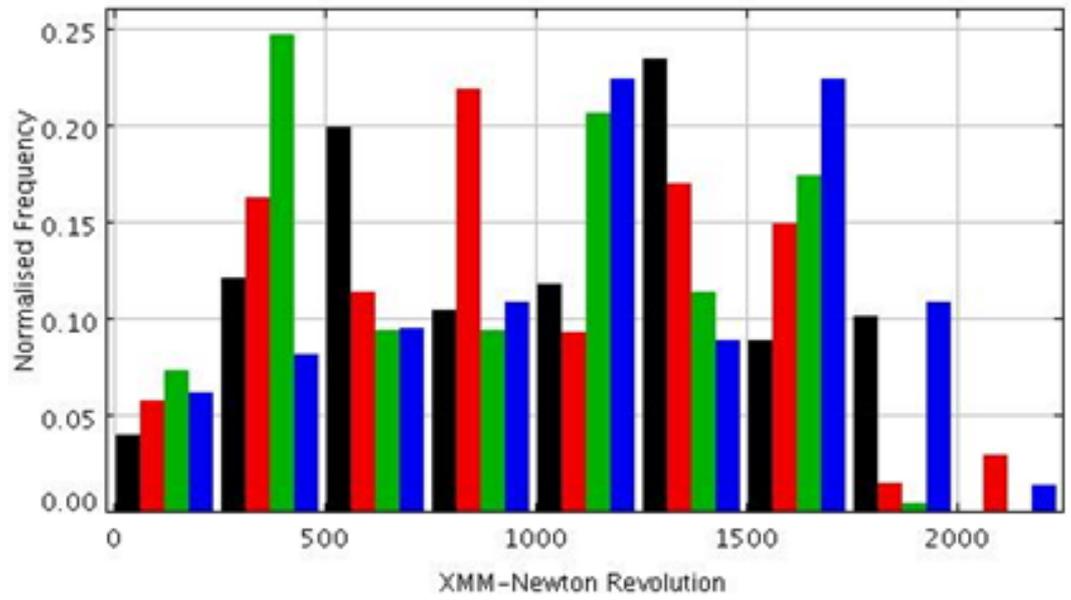

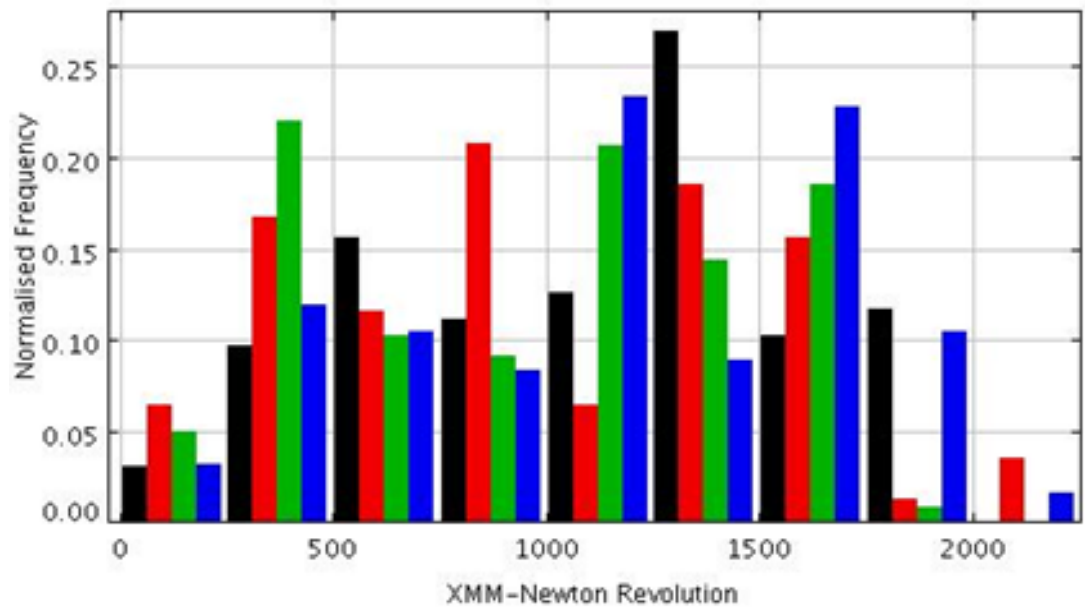

**Figs. 12a (top) and 12b.**

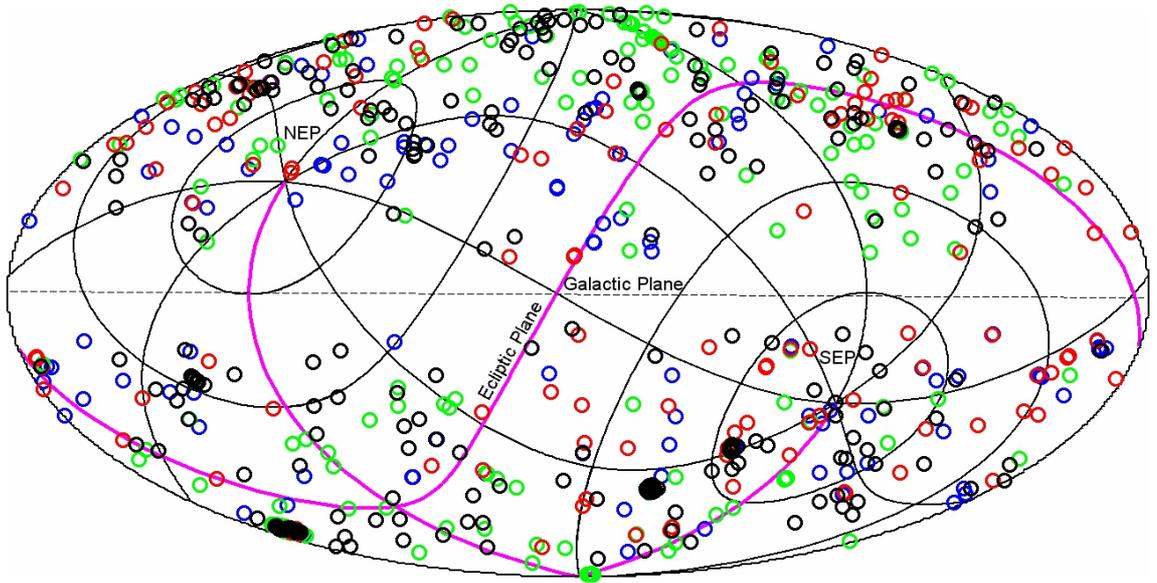
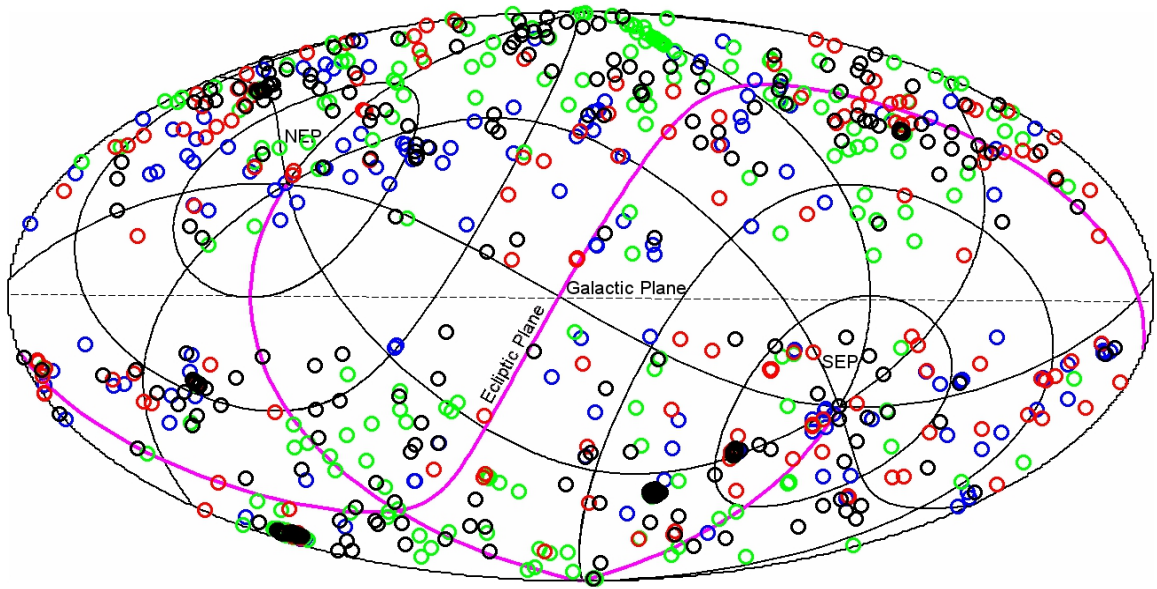

**Figs. 13a (top) and 13b.**

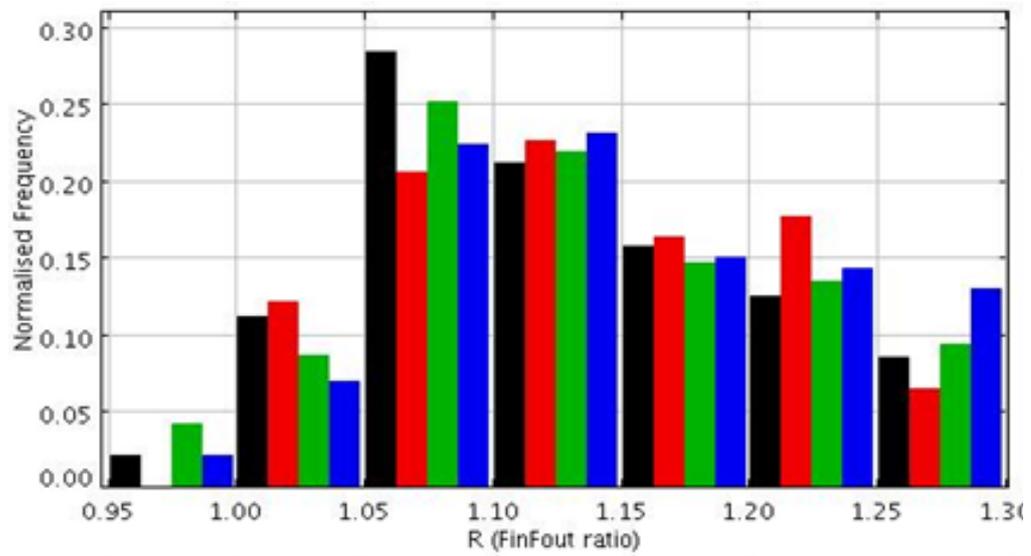
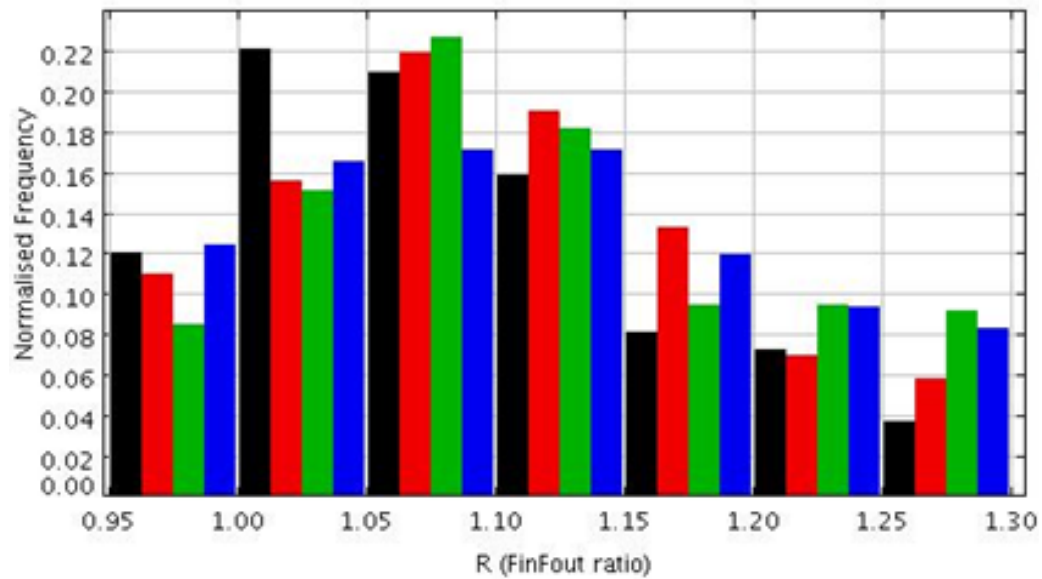

**Figure 14a.**

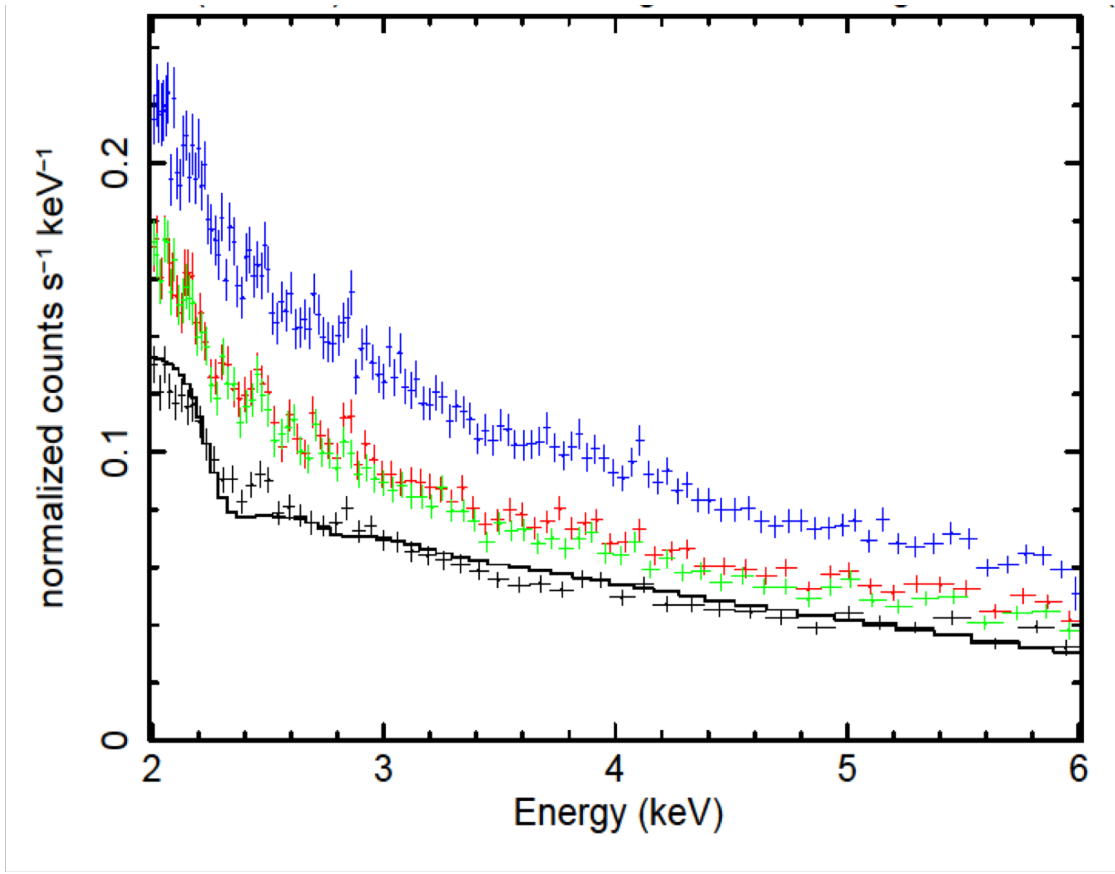

**Fig. 14b.**

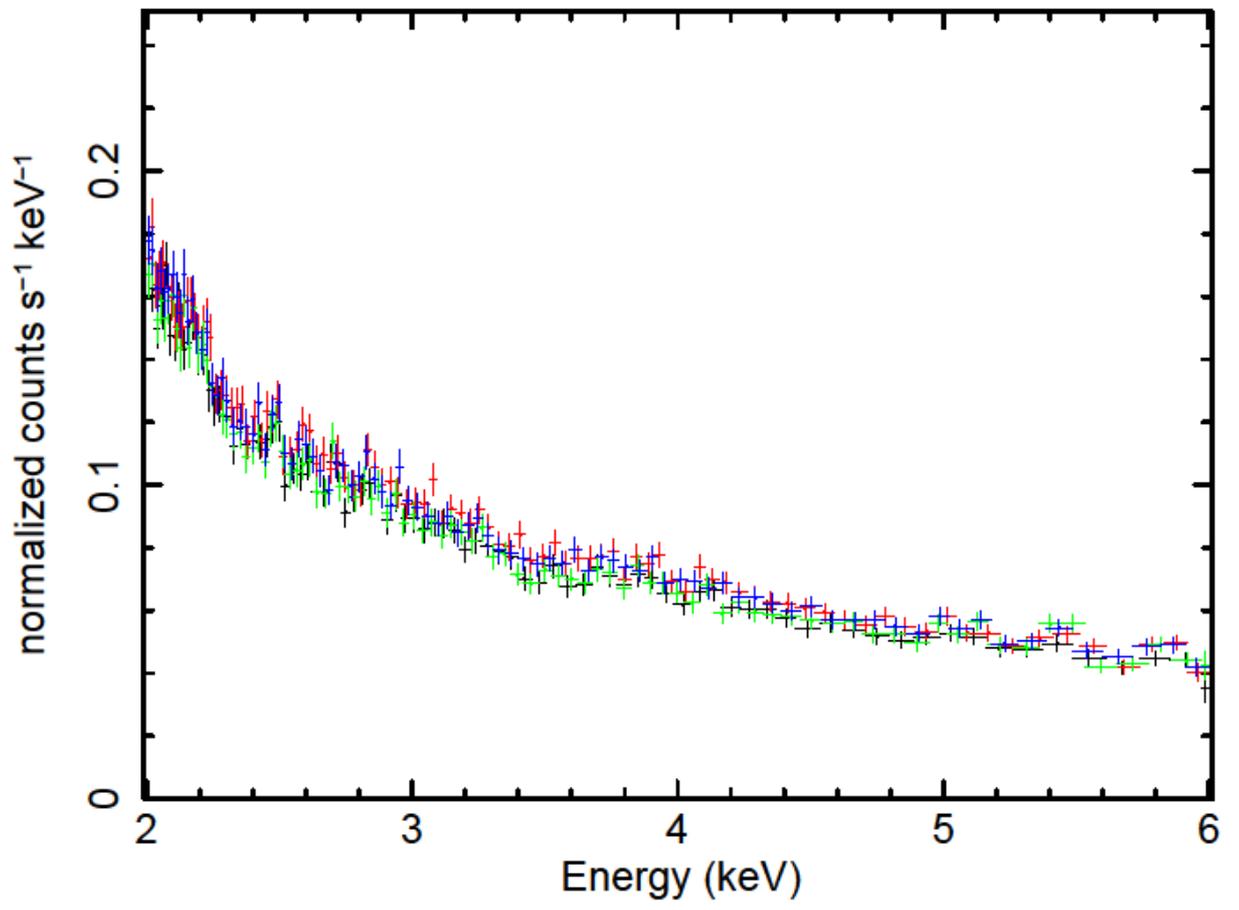

**Fig. 15a.**

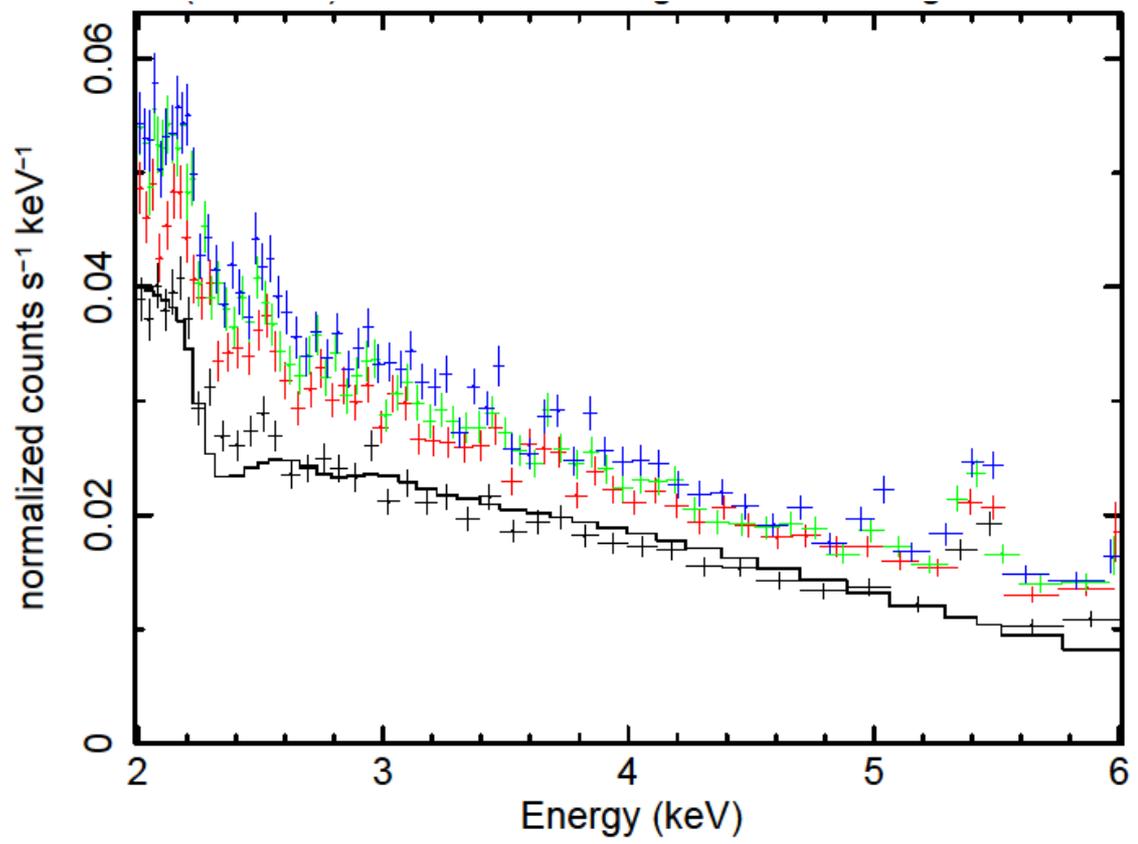

**Fig. 15 b.**

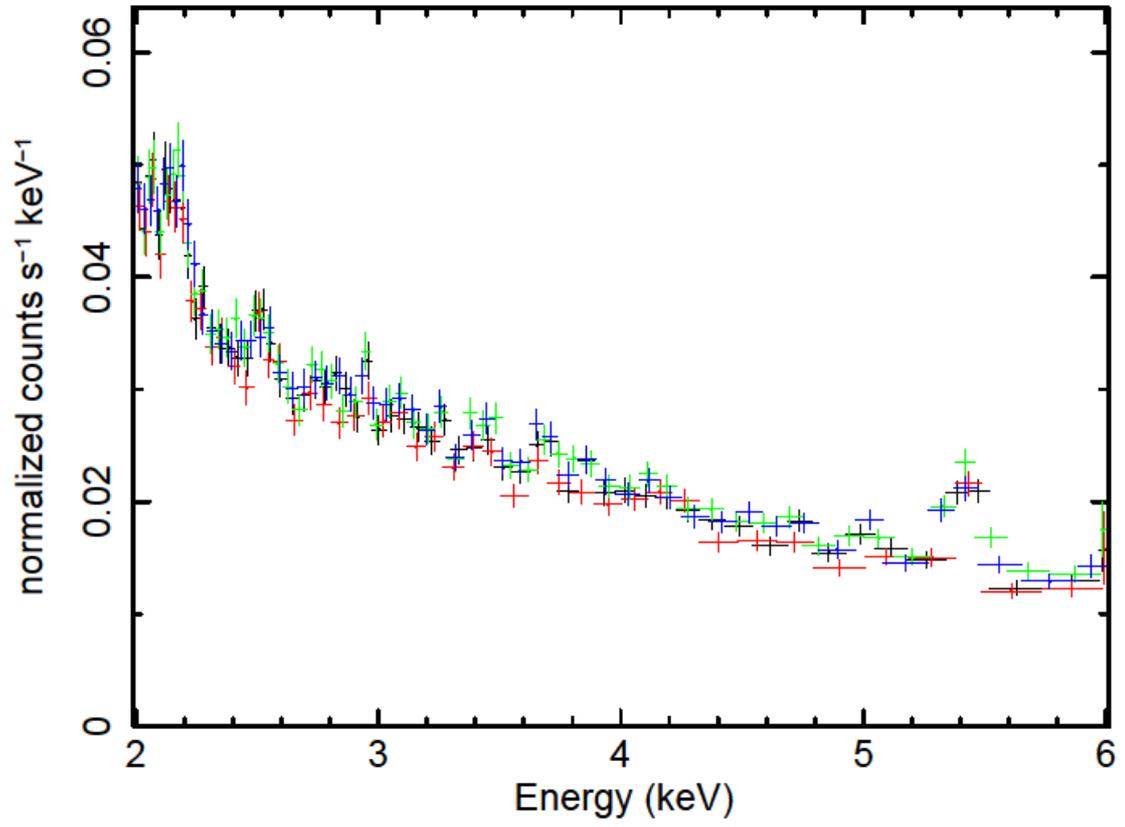

**Fig. 16.**

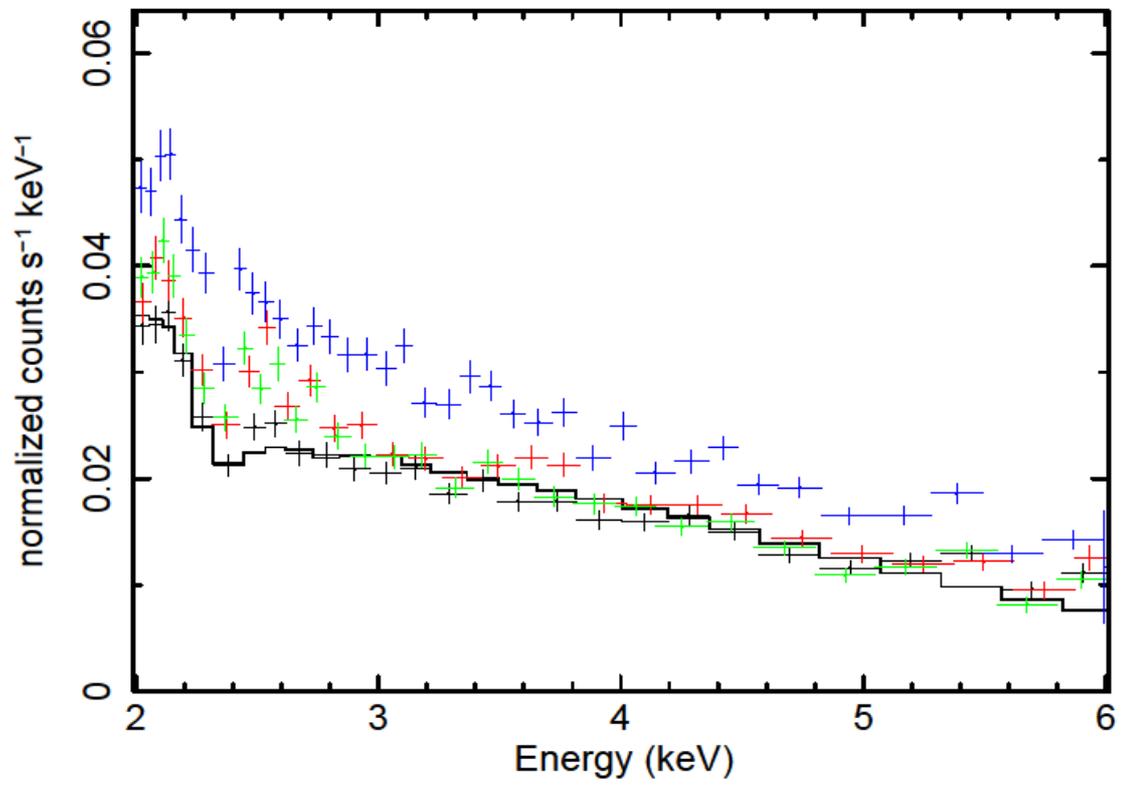

**Fig. 17a**

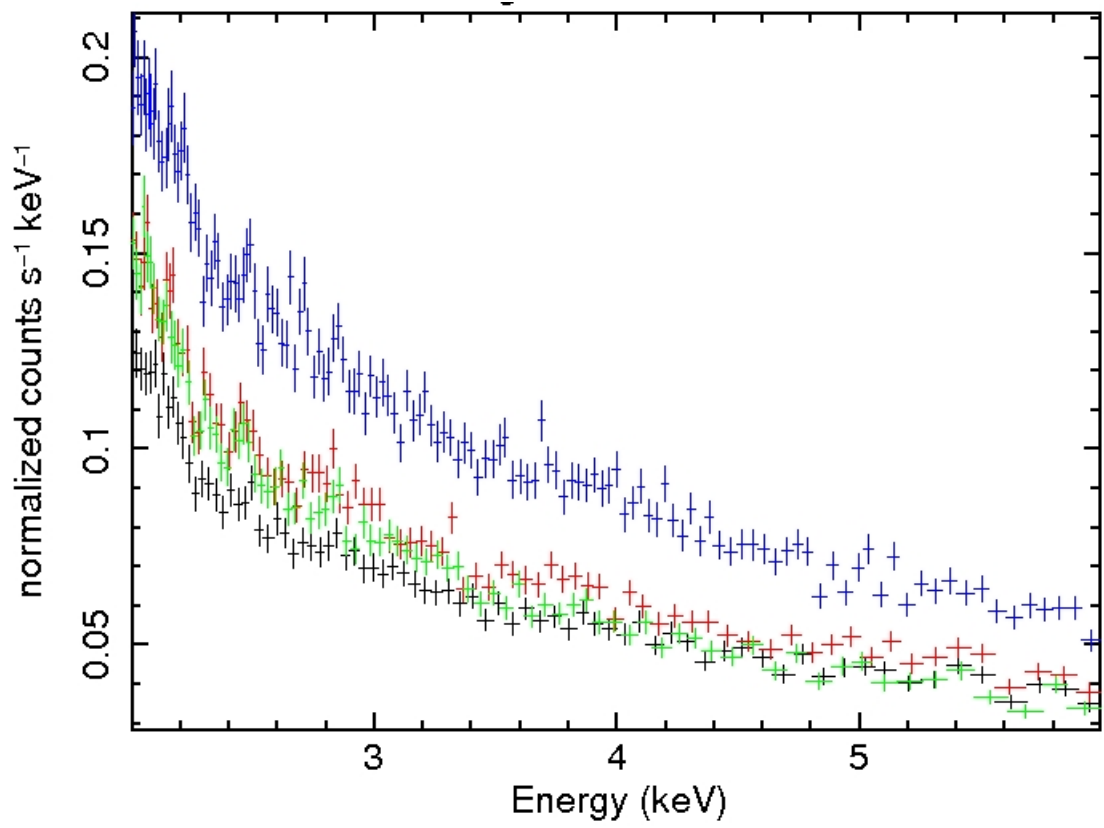

**Fig. 17b**

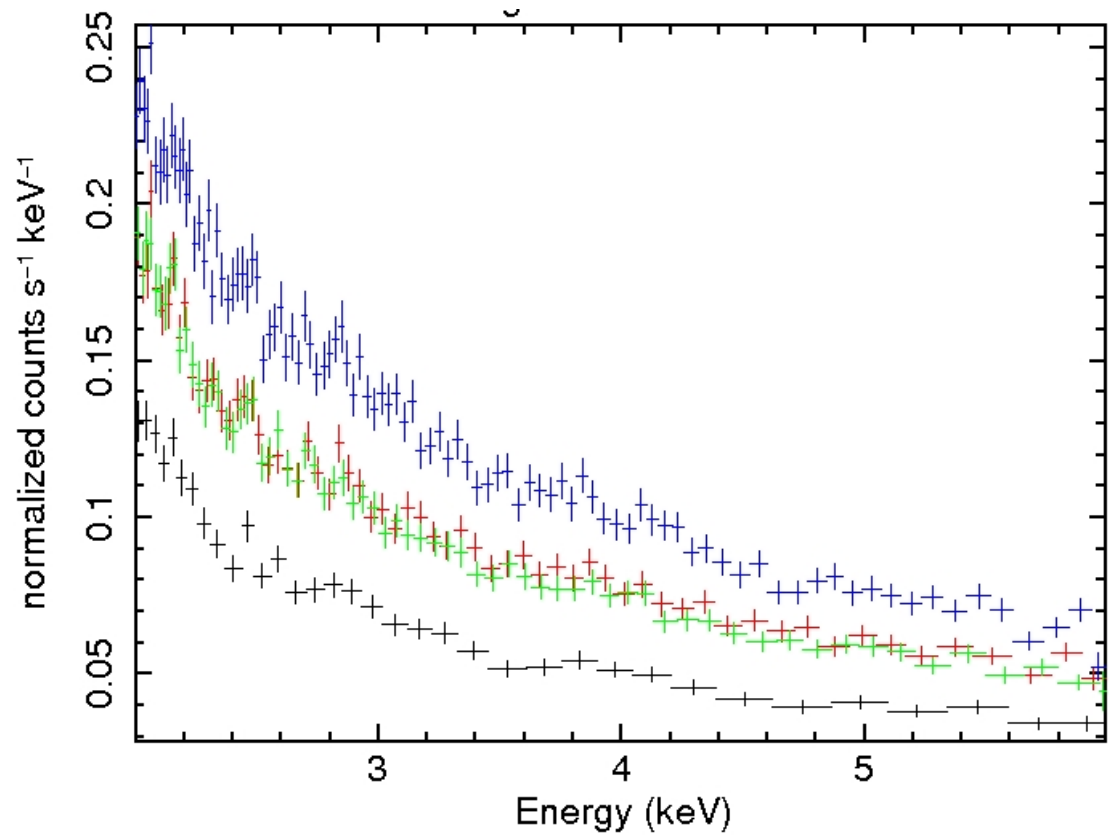

**Fig. 18.**

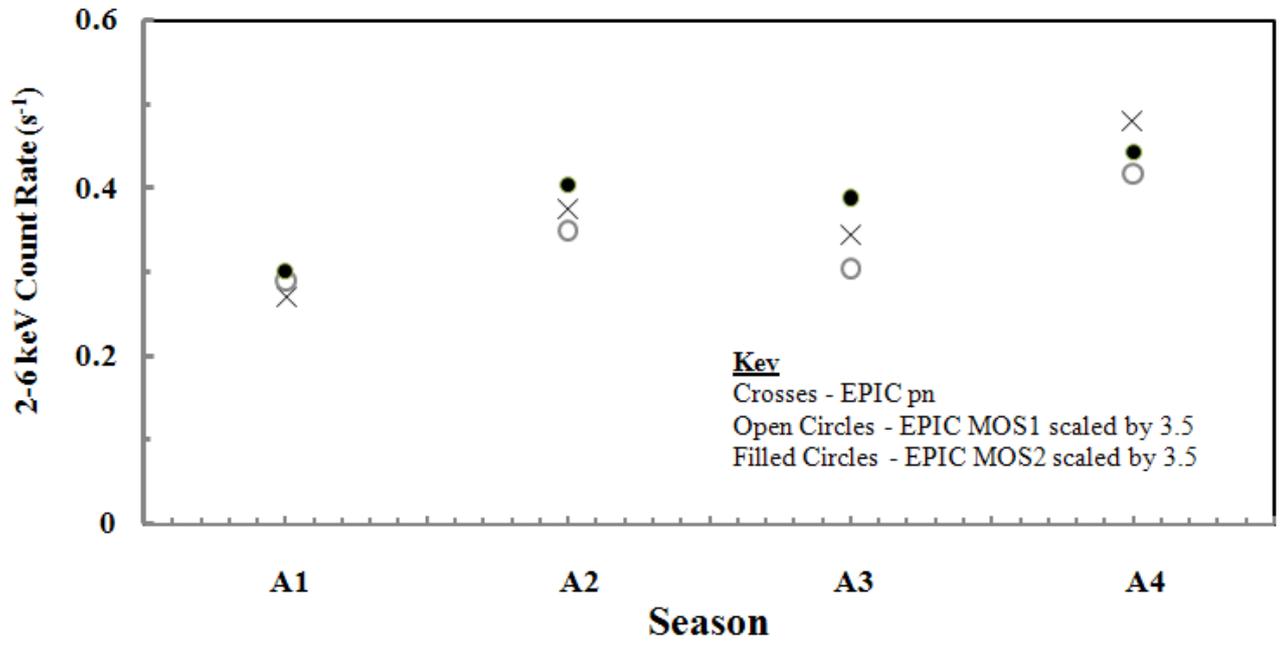

**Fig. 19.**

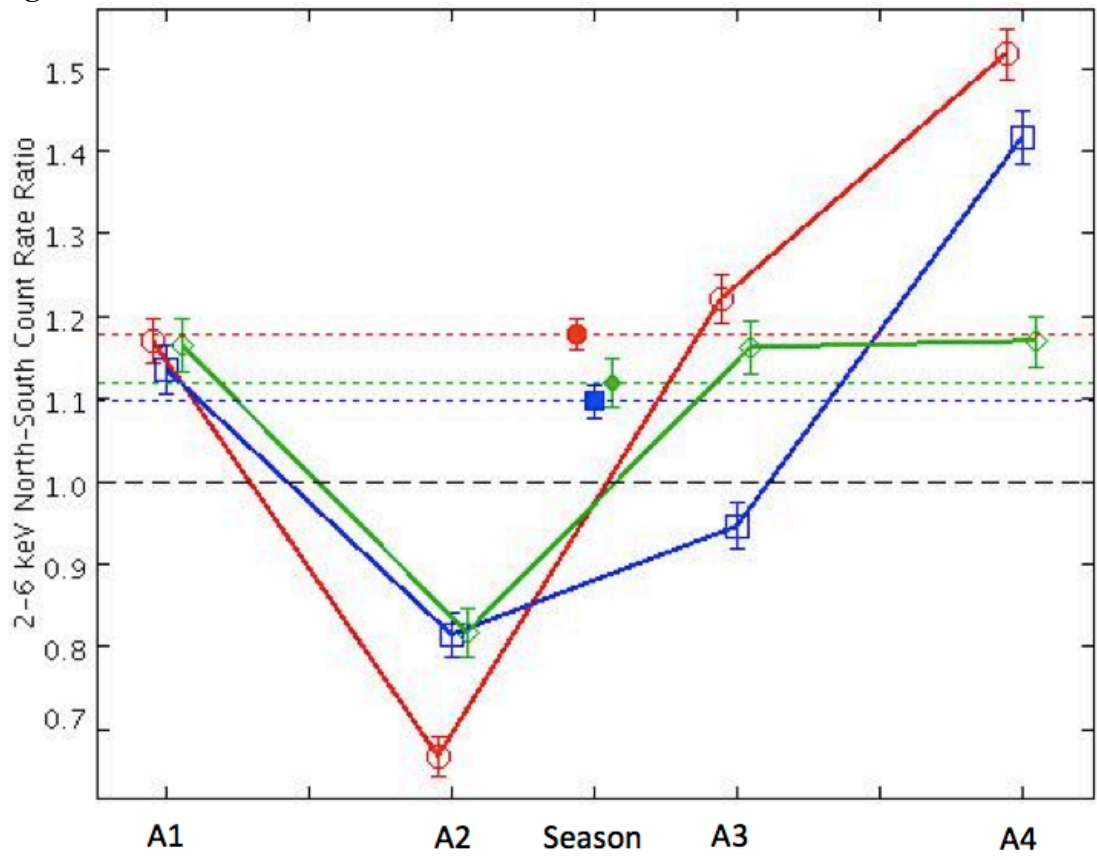

**Fig. 20.**

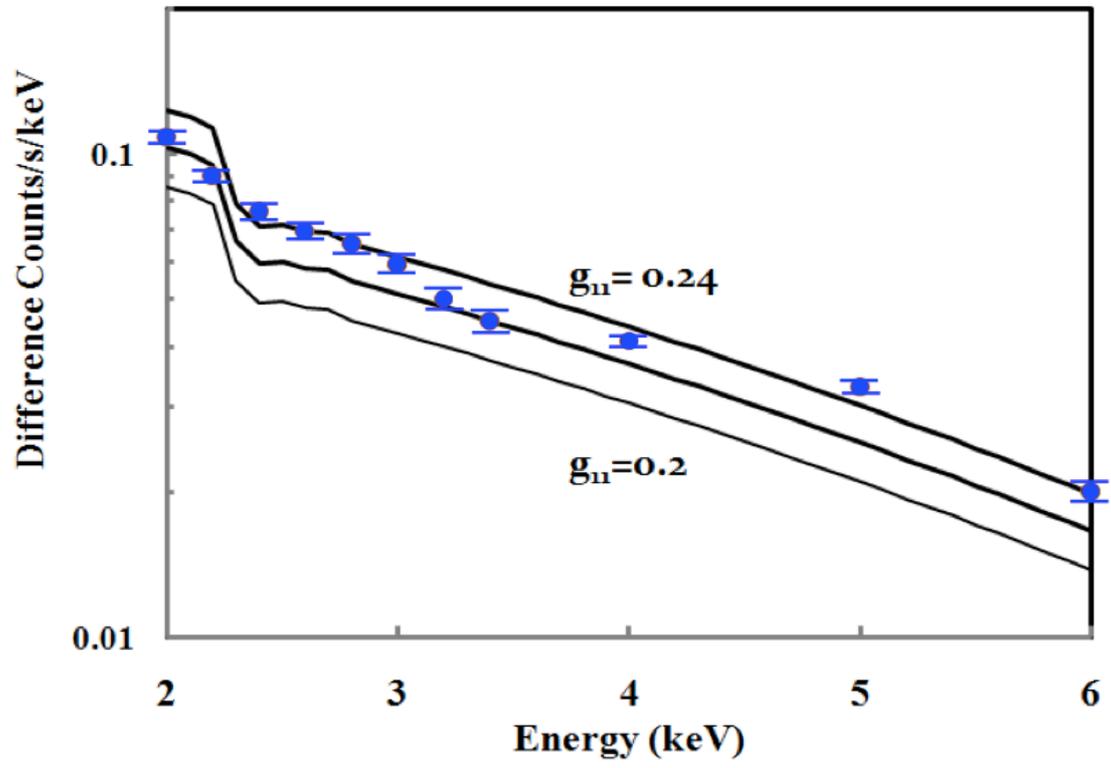

**Fig. 21.**

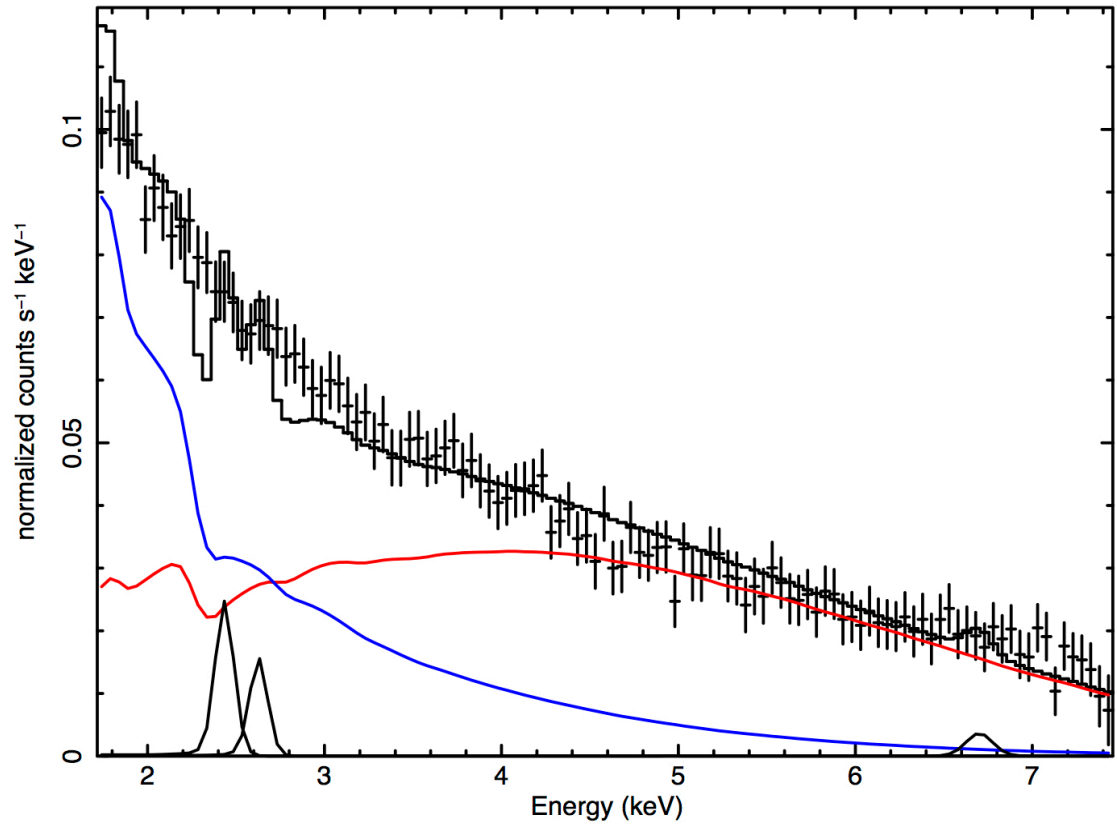

# APPENDIX

# Soft protons and the variable X-ray background

## 1. Introduction

Even after fourteen years of XMM-Newton operations, the interaction of clouds of magnetospherically-trapped 1-500 keV protons with the EPIC cameras has not been fully characterised, either analytically (using the modified Landau model of charged particle energy deposition in silicon CCDs described by Owens and McCarthy (1995)) or using the GEANT4 Monte Carlo package (Tenzer et al. 2008).

There are also fundamental difficulties in reconciling a seasonally variable background component – which appears coherent over more than one complete eleven-year solar cycle – to a background mechanism whose defining characteristic is short-term unpredictability. Any trapped particle model must also cope with the excursions of XMM-Newton's orbit beyond the magnetosphere into interplanetary space (see Appendix Fig. A1) and with the effects of varying pointing direction and orbital velocity on the proton number swept out per unit time. Nevertheless, in order to conclusively identify the X-ray signature of solar axion conversion, we must find one or more features to distinguish that signal from a "default" quiescent soft proton background in the EPIC pn and EPIC MOS cameras.

Alongside a warning that the SP spectrum in EPIC MOS below 2 keV is "uncharacterisable" (Kuntz & Snowden 2010), the literature assigns the following generic properties (Snowden Collier & Kuntz 2004; Kuntz & Snowden 2008) to the SP background:

**(a)** The quiescent SP spectrum is flatter than the canonical cosmic X-ray background (CXB) spectrum i.e. the exponent $\Gamma$ in a power law representation of the flux $F^{SP} = kE^{-\Gamma}$ is less than 1.4. The higher the SP intensity, the smaller the value of $\Gamma$.

**(b)** There is a spectral break at about 3.2 keV.

**(c)** The SP spectrum is expected to be quite featureless in the vicinity of the gold M edges since proton reflection from the gold-coated mirror shells should not "see" the absorption edges as soft X-rays do.

There is, however, no firm agreement as to the actual mechanism(s) underlying proton reflection. The multiple-scattering models developed immediately before and after the launch of XMM-Newton (Nartallo et al. 2001; Lo & Srour 2003) have been subsequently refined to include quasi-specular (Firsov) scattering (Lei et al. 2004) in order to match proton reflectivity measurements made on representative mirror flats. Aschenbach (2007) has used a quite distinct de Broglie wave analysis to estimate the proton grasps of EPIC pn, EPIC MOS and RGS.

**(d)** The vignetting function for soft protons decreases less rapidly with increasing off-axis angle than does the vignetting function for X-rays (Kuntz & Snowden 2008).

Observations (a), (b) and (c) do not greatly constrain the fitting of SP spectra, while (d) is puzzling. The XMM-Newton optics are very tightly baffled. If the soft protons were following exactly the same ray paths through the optics as the X-rays photons of the diffuse cosmic X-ray background, then, naively, the two vignetting functions should be the same.

The original scattering analysis of Nartallo et al. (2001) found that relatively few soft protons were reflected solely from the paraboloid and hyperboloid mirror shells, but arrived in the focal planes via a combination of surfaces, including baffles. In the case of RGS, 80% of soft protons reaching the detector plane had not scattered at all, from any surface. Thus, the RGS background count rate appeared to be the best available measure of the SP flux and of its seasonal variation. This conclusion is now questionable, since the original Nartallo model has been superceded (Lei et al. 2004).

## 2. Two-component Soft Proton Interaction Model
## 2.1 Direct Detection

The directly-detected SP count rate may be written in general terms, as a function of equivalent X-ray energy $E_x$ :

$$N_x^{SP}(E_x) = F^{SP}(E_p)G^{SP}(E_p)Q_p(E_p,E_x) \qquad \text{-(A1)}$$

The energies ($E_p$) and fluxes ($F_p$) of the proton populations in the Earth's magnetosphere have been assessed by a number of authors. Aschenbach (2007) has calculated the proton grasp-versus-proton energy functions $G^{SP}$ for all three of XMM-Newton's detection channels - EPIC pn, EPIC MOS and RGS. The EPIC focal plane cameras differ by the familiar aperture factor ~2, with the proton grasp of the RGS lower still by an order of magnitude, because of the effect of a third grazing-incidence reflection on the proton transmission. In all three cases, the proton grasp falls very sharply with increasing proton energy above about 400 keV and is constant at very low energies.

$Q_p$, the final term in Appendix eq. A1, describes the relationship of the incident proton energy to the energy deposited in the active silicon volume of the focal plane detector.

The range $R_p$ and stopping power $dE_p/dx$ of an energetic proton in any detector layer can be calculated using the National Institute of Standards and Technology (NIST) online utility PSTAR [12]. The range of a 100 keV proton in silicon is about 1 μm. Softer protons have ranges much smaller than the pixel sizes (40 μm (MOS) or 150 μm (pn)) or active detector thicknesses (40 μm (MOS) or 280 μm (pn)) in the EPIC cameras and comparable with the thicknesses of the EPIC bandpass filters (Appendix Table A1) and the thicknesses of the dead layers at the CCD entrance surfaces. Here, we find an important difference between the RGS and EPIC pn detectors – both with a back-illuminated CCD geometry (i.e. electrodes on the exit side of the silicon wafer, opposite the surface of X-ray incidence) - and the front-illuminated CCDs of EPIC MOS, whose open electrode structure presents an additional barrier to soft protons (Hiraga et al. 2001). In the case of the ACIS camera on Chandra, these geometries are explicitly present in the CCD nomenclature; ACIS consists of both front illuminated (FI) and back-illuminated (BI) chips (Lo & Srour 2003).

The precise details of the EPIC MOS electrode design remain proprietary, but the differences in thickness between the open electrode (covering 40% of the pixel area) and the normal electrodes have been derived using a so-called "mesh experiment"(Hiraga et al. 2001). Using published values of the EPIC MOS quantum detection efficiency at 1.5 and 3 keV (i.e. above and below the silicon K absorption edge) and a simple X-ray absorption model, one can, however, readily derive an effective open electrode thickness - equivalent to ~0.1 μm silicon plus 0.15 μm polysilicon (i.e. amorphous silicon dioxide). Over the remainder of the pixel area, the X-ray flux is attenuated by about 0.3 μm silicon plus 0.75 μm silicon dioxide. The low-energy X-ray and proton responses of EPIC pn are limited only by a passivation layer, around 0.1 μm thick.

---

[12] http://physics.nist.gov/PhysRefData/Star/Text/PSTAR,html

Since the proton stopping power $dE_p/dx$ has a broad maximum at about 100 keV in the materials of interest here (see Appendix Table A2), the energy loss in a given filter or detector layer may be treated as independent of proton energy- so that the transformation between $E_p$ and $E_x$ becomes, to a good approximation:

$$E_x = E_p - \Delta E(mirror) - \Delta E(filter) - \Delta E(electrode) = E_p - \sum \Delta E \qquad \text{-(A2)}$$

Then Appendix eq. A1 can be rewritten:

$$N_x^{SP}(E_x) = F^{SP}(E_x + \sum \Delta E) G^{SP}(E_p) \qquad \text{-(A3)}$$

If a closed functional form can be found for $F^{SP}$, and the energy loss terms $\Delta E$ are known, the X-ray equivalent spectrum due to the arrival of the scattered or reflected protons in the focal plane can be estimated.

## 2.2   PIXE

The observed SP background spectrum results from the modification, via multiple energy loss processes, of the source proton flux entering the telescope aperture – plus, inevitably, a second component due to proton-induced X-ray emission (PIXE) in the bandpass filters which precede the CCD focal plane detectors in the optical path. That the optical blocking filters in the Chandra ACIS CCD camera (0.03μm Al + 0.2 μm polyimide + 0.1 μm Al) are a barrier to low-energy protons (i.e. $E_p$ < 100 keV) is recognised (Lo & Srour 2003), but the production of continuum X-rays in such filters does not appear to have been considered in the literature.

## 3.   Calculations
### 3.1   Filter energy loss and X-ray production

Appendix Table A1 describes the composition of the EPIC bandpass filters.

From the stopping power data of Appendix Table A2, we estimate the energy loss $\Delta E(filter)$ in a 55 nm thick Al layer to be ~ 6 keV and in a 45 nm layer of Sn, ~7 keV. For the EPIC Thick filter, the total energy loss is then ~44 keV, dominated by the polypropylene. The detailed GEANT4 simulations of Fioretti (2011) indicate an energy loss of ~35 keV for the same filter. For the Thin and Medium filters, our estimates of the total energy loss are ~28 keV and ~32 keV, respectively.

Since the proton range generally exceeds the thickness of the filter layer, the simplifying approximation generally used to compute PIXE yields - that X-ray production is uniformly distributed throughout the filter layer - is a good one. The microscopic processes of X-ray production by protons in thin films are described by Ishii (1995).

For X-ray energies $T_{lim} > E_x > T_m$, where:

$$T_m = [4M_e/M_p]E_p \qquad \text{-(A4)}$$

the dominant production process is atomic bremsstrahlung (AB). Here, $M_e$ and $M_p$ are the electron and proton masses, respectively, and $E_p$ is the proton kinetic energy. For 100 keV protons, the lower threshold energy is 0.22 keV, independent of filter composition. The upper energy limit $T_{lim}$ is given by a complex, material-dependent function (eq. 11 of Ishii (1995)).

**Table A1**
Composition of the EPIC filters.

| Filter Description | Composition |
|---|---|
| Thin | 40 nm Al + 160 nm polyimide |
| Medium | 80 nm Al + 160 nm polyimide |
| Thick | 45 nm Sn + 55 nm Al + 330 nm polypropylene |

**Table A2**
Stopping power versus energy functions for soft protons from PSTAR in filter and detector materials of given chemical composition and bulk density $\rho$. The stopping powers are given in units of keV micron$^{-1}$.

| Energy $E_p$ (keV) | Polypropylene $C_3H_6$ $\rho = 0.9$ g cm$^{-3}$ | Polyimide $C_{22}H_{10}N_2O_5$ $\rho = 1.43$ g cm$^{-3}$ | Silicon Si $\rho = 2.3$ g cm$^{-3}$ | Aluminium Al $\rho = 2.7$ g cm$^{-3}$ | Tin Sn $\rho = 7.29$ g cm$^{-3}$ |
|---|---|---|---|---|---|
| 10 | 63 | 66 | 77 | 80 | 76 |
| 20 | 75 | 81 | 101 | 104 | 102 |
| 50 | 99 | 106 | 126 | 128 | 143 |
| 100 | 101 | 111 | 116 | 121 | 163 |
| 200 | 78 | 90 | 89 | 87 | 143 |
| 400 | 49 | 61 | 66 | 77 | 102 |

Appendix Fig. A2 shows the X-ray yields from the EPIC Thin and Medium filters calculated from the AB formulae of Pascher and Miraglia (1990) for proton energies in the range 50-300 keV. The X-ray yields increase very rapidly with decreasing X-ray energy and with increasing input proton energy.

### 3.2 Soft proton vignetting

The generation of X-rays in a filter layer situated very close to the EPIC focal planes provides, at least in principle, a possible explanation for the claimed slow roll-off of the SP vignetting function. We note, however, that the vignetting function of an X-ray telescope illuminated by a truly diffuse flux is not exactly represented by any sequence of off-axis observations of a point source. In the latter case, there is a unique grazing angle associated with each point on each mirror shell. In the former case, X-rays arrive at every point in the aperture over a range of angles. A simple one-dimensional model of a nested Wolter Type 1 telescope suggests that, in fact, the fall-off in intensity with increasing off-axis angle is less for *any* diffuse source than for a point source.

In a one-dimensional model telescope made up of $N$ co-axial shells, with radii $r_m$ $(1 \leq m \leq N)$, the vignetting function for a point source is:

$$V_{Point}(\theta) = \sum_1^N 2\pi r_m (r_m - r_{m-1}) [\frac{(R(\alpha_m + \theta/4) + R(\alpha_m - \theta/4))^2}{2}] \quad \text{- (A5)}$$

$\theta$ denotes the off-axis angle and $\alpha_m$ is the angle between the $m^{th}$ mirror shell and the telescope's optical axis. $R$ is the reflectivity for unpolarised X-rays. The energy dependence of this vignetting function is obtained by only counting those shells for which the indicated values of grazing angle are less than the critical angle of reflection for gold. In order to estimate the equivalent function for a uniform, diffuse flux, the average of the extremal reflectivities is replaced by the continuous average over the same angular range.

Appendix Fig. A3 shows the relative response $V_{point}(\theta)/V_{point}(0)$ calculated for a point source of 2 keV X-rays incident upon a 15 shell approximation to the 60 cm diameter, 8.5m focal length XMM-Newton mirror assembly. For comparison, the calculation is repeated for a diffuse source of 2 keV X-rays. The GEANT4 estimate of the soft proton vignetting function provided by Fioretti (2011) closely follows our diffuse X-ray curve, falling to about 75% of the on-axis effective area at the edge of the telescope field-of-view.

### 3.3 Directly detected proton count rates

In the absence of any sensitive on-board radiation monitor, XMM-Newton's ambient proton environment must be represented by contemporary instruments in deep space – such as the Advanced Composition Explorer (ACE) at the L1 Lagrange point[13] – and by derived particle flux models such as AP-8[14].

Remarkably, the proton records of the four identical Cluster spacecraft do not appear to have been previously used to estimate $F^{SP}$ for XMM-Newton, despite some similarities in orbital geometry. The soft proton fluxes recorded by Cluster in the equatorial plasma sheet region ($-3R_E < z < 3R_E$) of the nightside magnetosphere, extending in energy from 1 eV to 1 MeV, measured under differing solar conditions –from quiet Sun to C, M and X-flare states – from dawn round to dusk and from July to October (when the Cluster spacecraft are preferentially in the magnetotail) have all been represented by a simple Kappa function (Haalaand et al. 2010):

$$F^{SP}(E_p) = A(E_p/E_0)[1 + \frac{E_p}{kE_0}]^{-k-1} \qquad \text{- (A6)}$$

---

[13] The ACE payload is described in a dedicated issue of Space Science Reviews - 86(1-4) 1998.
[14] http://www.spenvis.oma.be

A is a normalisation constant. The parameters $k$ and $E_0$ appropriate to low levels of magnetospheric activity are:

$$2 < k < 5 \qquad 2\text{ keV} < E_0 < 4\text{ keV}$$

Combining Appendix Eqs. A2, A3 and A6, we can now estimate the analytical form of the quiescent SP spectrum.

For EPIC pn and the Thick filter the total energy loss $E_p - E_x$ is 55 keV, assuming 3 keV proton energy loss per Firsov reflection and a loss of 5 keV in the pn passivation layer,

For EPIC MOS and the Thick filter, there are two possible outcomes. If the proton strikes the open fraction of the CCD pixel, the energy loss is the same as for EPIC pn, plus about 20 keV. If the proton has to penetrate the 1.05 μm equivalent silicon of the thicker electrodes, the total energy loss is much higher – not 55 keV, but 155 keV. Then for an X-ray equivalent energy of 2 keV, the ratio of count rates is expected to be:

$$[N_x^{SP}]_{pn}/[N_x^{SP}]_{MOS} = 2F^{SP}(57keV)/[0.4F^{SP}(77keV) + 0.6F^{SP}(157keV)] \qquad \text{- (A7a)}$$

where the prefactor 2 on the right hand side represents the ratio of proton apertures calculated by Aschenbach (2007). The corresponding equation for observations with the Medium filter is:

$$[N_x^{SP}]_{pn}/[N_x^{SP}]_{MOS} = 2F^{SP}(45keV)/[0.4F^{SP}(65keV) + 0.6F^{SP}(145keV)] \qquad \text{- (A7b)}$$

and for the Thin filter :

$$[N_x^{SP}]_{pn}/[N_x^{SP}]_{MOS} = 2F^{SP}(41keV)/[0.4F^{SP}(61keV) + 0.6F^{SP}(141keV)] \qquad \text{- (A7c)}$$

A practical lower limit to the ratio of SP count rates in the two detectors is given by the ratio of their proton apertures divided by the open electrode fraction of the MOS CCDs. That is:

$$[N_x^{SP}]_{pn}/[N_x^{SP}]_{MOS} = 2/0.4 = 5 \qquad \text{- (A7d)}$$

already significantly higher than the equivalent X-ray ratio derived in Section 2.3.2 of the main paper. Substituting $k = 2$ and $E_0 = 2$ keV in eq. 6 leads to a count rate ratio of 5.7 from eq. 7a, while eq. 7b yields a ratio of 6.24. These results are sensitive, in particular, to the energy loss value assumed for the EPIC pn passivation layer.

Fig. 6.20 of Fioretti (2011) compares simultaneous EPIC pn and EPIC MOS observations of a SP flare using the Thick filter; the ratio of count rates in the 2-10 keV band is 20:1 rather than 5:1, but is inflated by the higher contribution of cosmic ray background events in the case of EPIC pn.

A stronger confirmation of eq. A7d is presented in Appendix Figs. A4a,b. The top panel (Fig. A4a) shows the 2-6 keV light curve of a well-studied EPIC pn full frame observation (Observation ID: 0085150301), the subject of a previous investigation of SWCX emission by three of us (Carter, Sembay & Read 2010). The X-ray signal and charged particle background within this observation are well characterised and the obvious flaring seen in the lightcurve is certainly due to soft protons. The bottom panel (Fig. A4b) shows the background subtracted EPIC pn count rate plotted against the background subtracted EPIC MOS1 count rate from the same time bins. Error bars have been excluded for clarity except for the point with the highest observed rates. The EPIC pn background rate was determined from the quiescent period bounded by the two dashed lines in Fig. A4a. All valid event pattern types were selected; the same region was used within the field of view of both cameras. A small correction to the EPIC pn count rate was made to account for the small differences in active area (due to CCD gaps and bad-pixel subtraction) between the cameras.

The respective soft proton count rates are highly correlated, with the possible exception of the bin with the very highest rate. The slope of the pn-to-MOS1 graph is 5.13, in very good agreement with the theoretical arguments developed above. The gradient derived from the corresponding EPIC MOS2 data set (not shown) is 5.19.

Fig. A5, finally, draws together the elements of this Appendix in the form of a calculated, two-component quiescent SP spectrum for EPIC pn. Above 3 keV, the spectrum is essentially flat and the predicted count rates are at least one order of magnitude below the A4-A1 difference count rates of Fig. 18 in the main paper.

**Fig. A1**

Radial distance of the XMM-Newton spacecraft from the Sun-Earth line (solid lines) compared with the extent of the Earth's magnetosphere (broken lines). GSE coordinates are used, and a parabolic magnetosphere model with a stand-off distance of 10 $R_E$. The time axis extends for one complete revolution. Red lines show the winter orbital geometry, for 1$^{st}$ January 2000. Black lines show the summer geometry, for 1$^{st}$ July 2000. When the full curve lies above the broken curve, as it does for most of the Summer orbit, the spacecraft is formally outside the magnetosphere in interplanetary space and a trapped SP background mechanism is implausible.

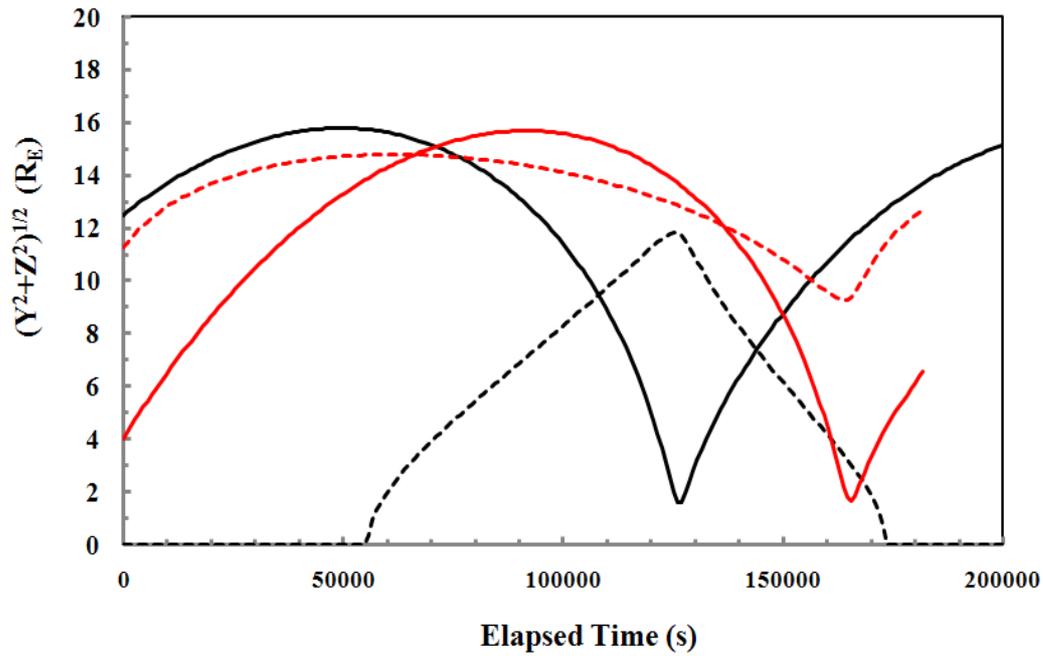

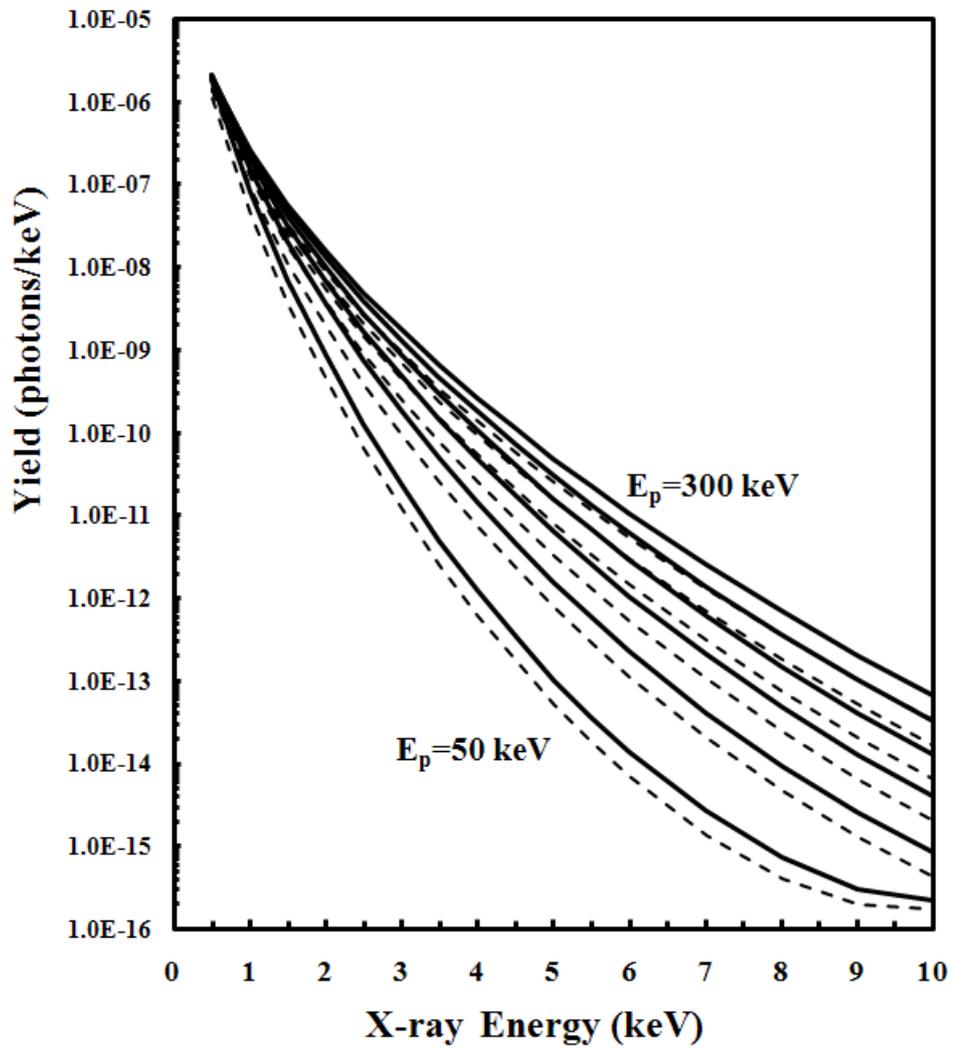

**Fig. A2.**

Calculated atomic bremsstrahlung yields from Thin (broken curves) and Medium (full curves) filters. Proton energies are spaced from 50 keV to 300 keV at 50 keV intervals.

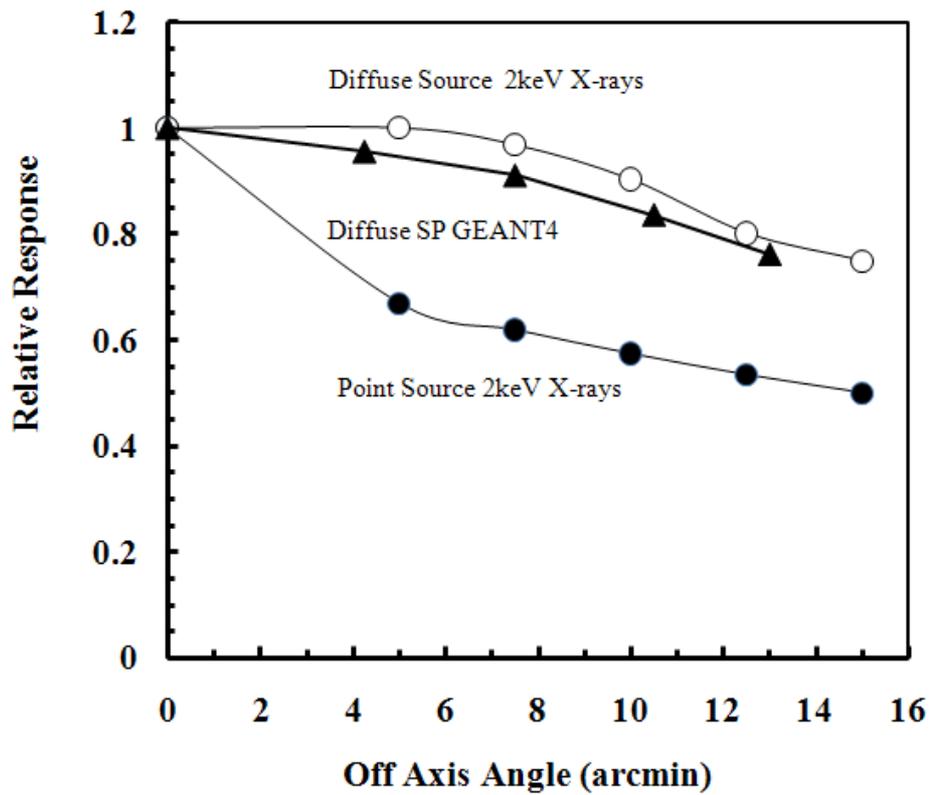

**Fig. A3.**

Comparison of the calculated vignetting functions for a 2 keV point source (filled circles) and a uniform diffuse source of the same energy (open circles). The two curves broadly reproduce the results presented by Snowden et al. (2004) for point X-ray sources and for soft protons but the difference between them lies in the geometry of the illumination and not in the nature of the radiation. The GEANT4 SP calculation of Fioretti (2011) is indicated by the triangles.

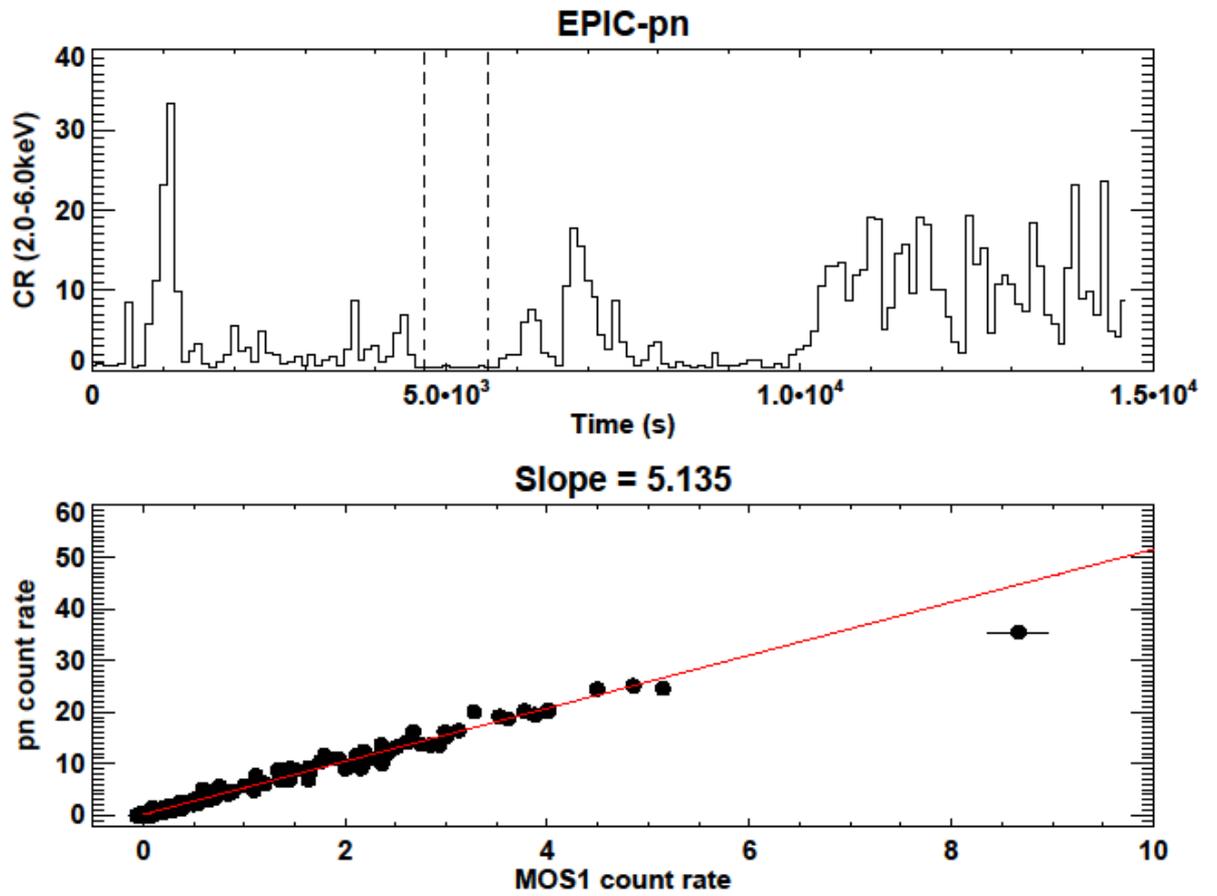

**Fig. A4**

**(a)** (Top panel) EPIC pn light curve for Obs. ID 0085150301 (Carter, Sembay & Read 2010), showing severe soft proton flaring. **(b)** (Bottom panel) relationship between instantaneous EPIC pn and EPIC MOS1 count rates.

**Fig. A5.**

Calculated two-component quiescent SP spectra for EPIC pn and Thick filter (full curves). PIXE component for Medium filter is indicated by the broken curve. The input spectrum is a kappa function with $E_0$ = 2 keV and $k$ = 2, and normalisation $10^5$ protons cm$^{-2}$ s$^{-1}$ keV$^{-1}$ sr$^{-1}$. The EPIC pn proton grasp is from from Aschenbach (2007). Broken vertical lines bound the energy region in which atomic bremsstralung is the dominant X-ray production mechanism (see Appendix Section 3.1).

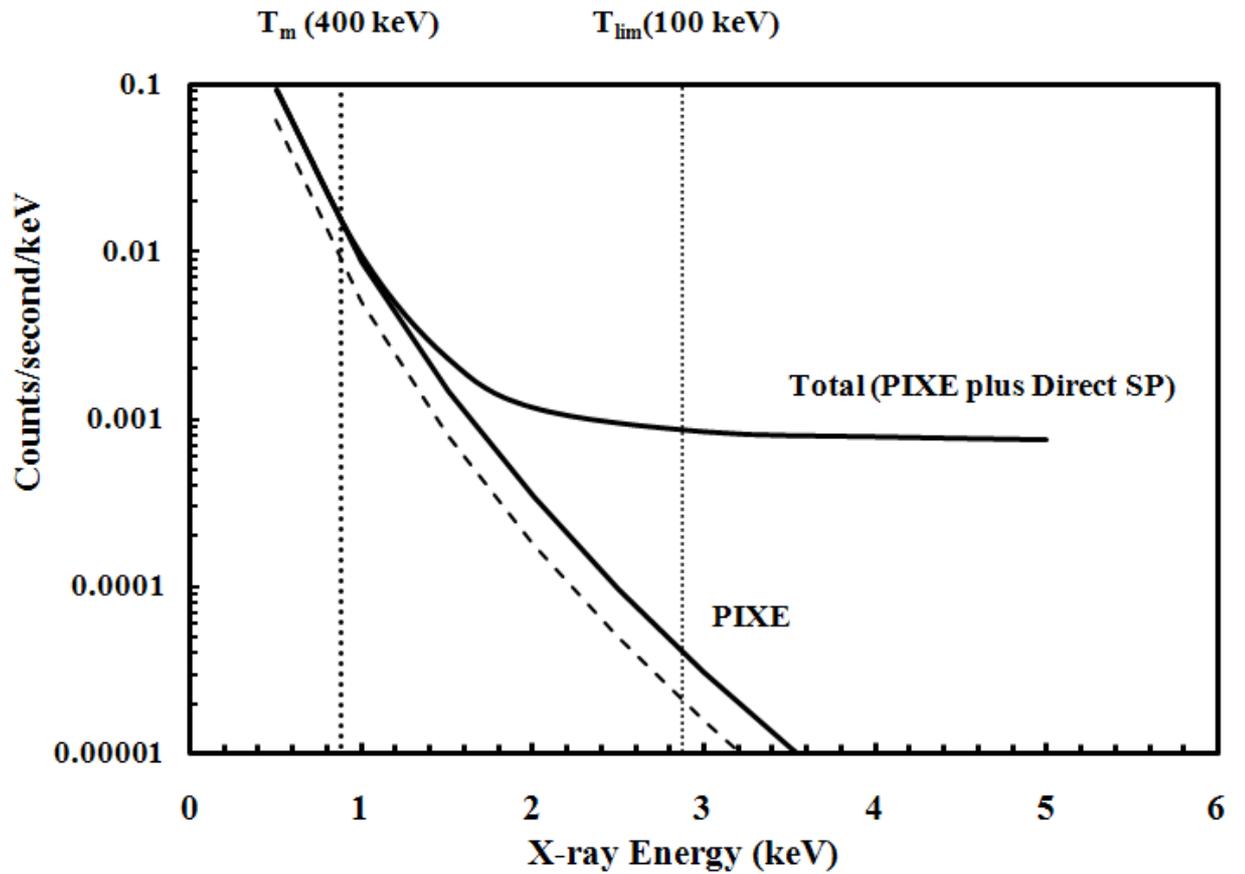